\RequirePackage{fix-cm}
\documentclass{svjour3} 
\smartqed 
\usepackage{graphicx}
\usepackage{bm}
\usepackage{epsfig}
\usepackage{color}

\include{psfig}

\journalname{Flow Turbulence Combust.}
\begin{document}

\title{Vertical motions of heavy inertial particles smaller than the smallest scale of the turbulence in strongly stratified turbulence
}

\titlerunning{Vertical motions of heavy inertial particles in strongly stratified turbulence}

\author{
{F. C. G. A. Nicolleau} \and {K.-S. Sung} \and {J.~C.~Vassilicos}
}

\institute{F. C. G. A. Nicolleau \at
The University of Sheffield, SFMG, Department of Mechanical Engineering,
Mapping Street, Sheffield, S1 3JD, United Kingdom \\
\email{F.Nicolleau@Sheffield.ac.uk} 
\and
K.-S. Sung \at
Imperial College London, Department of Aeronautics,
Prince Consort Road, SW7 2BY, United Kingdom
\and
J. C. Vassilicos \at
Imperial College London, Department of Aeronautics,
Prince Consort Road, SW7 2BY, United Kingdom \\
\email{J.C.Vassilicos@Imperial.ac.uk}
}

\date{Received: date / Accepted: date}

\maketitle

\begin{abstract}
We study the statistics of the
vertical motion of inertial particles in
strongly stratified turbulence.
We use Kinematic Simulation (KS) and Rapid Distortion Theory (RDT) to study the
mean position and the root mean square (rms) of the position fluctuation in the vertical direction.
We vary the strength of the stratification and the particle inertial characteristic time.
The stratification is modelled using the Boussinesq equation and solved in the limit of RDT.
The validity of the approximations used here requires that
$
\sqrt{{L}/{g}}
<
{2\pi}/{\mathcal{N}}
<
\tau_{\eta}
$,
where $\tau_{\eta}$ is the Kolmogorov time scale, $g$ the gravitational acceleration, $L$ the turbulence integral length scale and $\mathcal{N}$ the Brunt-V\"ais\"al\"a frequency. 
We introduce a drift Froude number
$Fr_{d} = \tau_p g / \mathcal{N} L$.
When $Fr_{d} < 1$, the rms of the inertial particle displacement fluctuation is the same as for fluid elements, i.e.
$\langle(\zeta_3 - \langle \zeta_3 \rangle)^2\rangle^{1/2} = 1.22\, u'/\mathcal{N} + \mbox{oscillations}$.
However, when $Fr_{d} > 1$,
$\langle(\zeta_3 - \langle \zeta_3 \rangle)^2\rangle^{1/2} = 267 \, u' \tau_p$. That is the level of the fluctuation is controlled by the particle inertia $\tau_p$ and
not by the buoyancy frequency $\mathcal{N}$. In other words it seems possible for inertial particles to retain the vertical capping while loosing the memory of the
Brunt-V\"ais\"al\"a frequency.



\end{abstract}

\maketitle

\section{Introduction}
\label{chp:1}

The vertical transport of inertial
particles in stably stratified turbulence is important in order to
understand the behaviour of heavy particles such as droplets in clouds,
dust or pollutants.
Stratification can be found in many geophysical or industrial flows
(e.g. diffusion of pollutants in the atmosphere or ocean, movement and growth
of clouds).
The term {\it stratified flow} is normally
used for {\it flow of stratified fluid}, or more precisely
{\it density stratified fluid}, and this is the meaning it has in this paper. In these fluids,
the density varies with the position in the fluid,
and this variation is important in term of fluid dynamics. Usually, this density
variation is stable with nearly horizontal lines of constant density, i.e. lighter fluid
above and heavier fluid below. The density variation may be
continuous, as it occurs in most of the atmosphere and oceans, this is the case we consider
in this paper, that is a fluid with a negative density gradient in the
vertical direction. In many situations the variation of density is very small.
However, this small variation can have a severe effect on the flow if the
small buoyancy forces can come into play. A stably stratified turbulence
has a vertical structure which is different from that of isotropic
turbulence, and which leads to vertical depletion of fluid particle diffusion
(see e.g. \cite{Kaneda-Ishida-2000,Nicolleau-Vassilicos-2000}).
\\[2ex]
The whole Eulerian field may be given by a Direct Numerical Simulation (DNS) as in
\cite{Aartrijk-Clercx-2009}
with the usual limitation in terms of Reynolds number and high computing cost.
In this paper we use a synthetic model of turbulence: Kinematic Simulation (KS), to study the statistics of the vertical motions of heavy particles in strongly stratified turbulence.
KS allows large Reynolds numbers and regimes which are not achievable with DNS. Focusing on asymptotic cases and monitoring the
construction of a synthetic field allows one to understand the respective role of Eulerian and Lagrangian correlations (see e.g. \cite{Cambon-al-2004,Nicolleau-Yu-2007}) and of the nolinear terms.

One particle diffusion\footnote{In this paper, particles are synonymous with
fluid elements, so particle diffusion means the dispersion of a fluid particle. This is by contrast to heavy particle or inertial particle.}
in stratified flows has already received much attention~\cite{kimura-Herring96,Godeferd-al-1997,Nicolleau-Vassilicos-2000,Kaneda-Ishida-2000}
and validation of KS has been made against DNS.%
There is less work devoted to {\it heavy particles} immersed in stratified flow.
In this paper, we consider such heavy particles, that is particles which are heavier than the surrounding fluid.
\\[2ex]
The lowest typical Froude numbers presented in experimental studies are larger than 0.01. Those Froude numbers can be reproduced by either DNS or our KS model
but only at low Reynolds numbers. On the one hand, DNS solve the Navier Stokes equations without assumptions but are far from reaching Reynolds numbers relevant to e.g. atmospheric or oceanic flows.
On the other hand, KS can easily achieve large Reynolds numbers but at the cost of the RDT assumption 
and for that reason are limited to lower Froude numbers. 
The larger Froude numbers used in our model here are $Fr\simeq 0.002$, (${\cal N}=500$).
The method presented here is a complement to DNS and experiments. 
Comparisons between DNS and KS at low Reynolds numbers help to understand the
respective role of linear and no-linear terms for stratified flows (e.g. \cite{Nicolleau-Yu-2007}). In this paper we extrapolate to flows with higher Reynolds numbers (not achievable with DNS) but at the cost of decreasing the Froude number which may be lower
than what is encountered for practical flows. For those high Reynolds flows, we discovered non-intuitive new regimes for heavy particle dispersion.

\section{Numerical Models}

\subsection{Kinematic Simulation (KS)}

The Kinematic Simulation technique (KS) was first developed
for incompressible isotropic turbulence
\cite{Fung-al-1992}. This model is based on a kinematically
simulated Eulerian velocity field which is generated as a sum of
random incompressible Fourier modes. This velocity field has a
turbulent-like flow structure, that is eddying, straining and
streaming regions, in every realization of the Eulerian velocity
field, and the Lagrangian statistics are obtained by integrating individual
particle trajectories in many realisations of this velocity field.

About a decade ago \cite{Godeferd-al-1997,Nicolleau-Vassilicos-2000}, KS was extended to anisotropic turbulence, specifically stably stratified homogeneous incompressible turbulence fluctuations. A step further was taken in 2004 \cite{Cambon-al-2004} when KS of stably stratified and/or rapidly rotating homogeneous and incompressible turbulence was discussed in detail as to its Lagrangian predictions. Here we use KS of stratified turbulence following \cite{Nicolleau-Vassilicos-2000}. We present this KS in the following subsection.

\subsection{Boussinesq approximation}

More details on KS and its use for one and two-particle diffusion
in stably stratified non-decaying turbulence can be found in
\cite{Godeferd-al-1997,Nicolleau-Vassilicos-2000,Cambon-al-2004,Nicolleau-Yu-2007,Nicolleau-al-2008}.
The KS model used here is based on the Boussinesq approximations.
A stably-stratified turbulence is given at static equilibrium, with
pressure $p(x_3)$ and density $\rho(x_3)$ varying only in the
vertical axis, that is the direction of stratification.
Hence, we have $dp/dx_3=-\rho g$ where $\mathbf{g}=(0,0,-g)$ is the
gravity.
For a stable stratification, the mean density
gradient is negative i.e. $d\rho/dx_3<0$ as the tilting of a density
surface will produce a restoring force. From the Boussinesq
approximation we have:
\begin{equation}
\frac{D}{Dt}\left(\frac{\rho'}{\rho}\right)=-u_3\frac{1}{\rho}\frac{d\rho}{dx_3}\label{eqn:density}
\end{equation}
where $D/Dt=\partial/\partial t+\mathbf{u}\cdot\nabla$ is the
Lagrangian derivative, $p'$ the perturbation pressure and $\rho'$ the
density fluctuation, this latter is much smaller than $\rho$
($\rho' \ll \rho$) so that, in the limit of a vanishing viscosity, the dynamic equation becomes:
\begin{equation}
\frac{D}{Dt}\mathbf{u}=-\frac{1}{\rho}\nabla
p'+\frac{\rho'}{\rho}\mathbf{g}
\label{eqn:lagstro}.
\end{equation}
The perturbation velocity $\mathbf{u}(x,t)=(u_1,u_2,u_3)$ is taken
incompressible
\begin{equation}
\nabla \cdot \mathbf{u}=0.
\label{eqn:incompre}
\end{equation}

\subsection{Linearized Boussinesq equations}

\begin{figure}[h]
 \includegraphics[width=11cm]{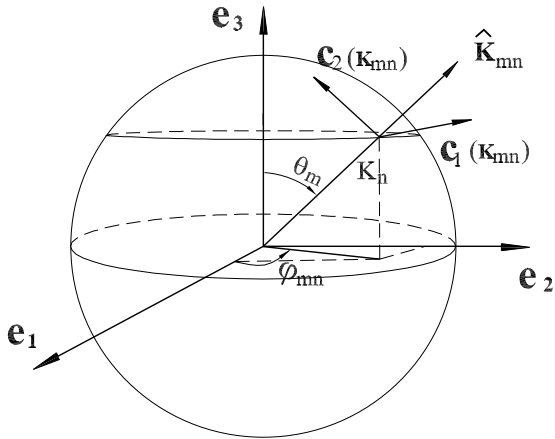}
\caption{Craya-Herring frame.}
\label{fig:craya}
\end{figure}

The initial velocity $\mathbf{u}(\mathbf{x},0)$ can involve a large range of length scales, the
smallest of these length scales is $\eta$, the Kolmogorov length scale. In the limit where
nonlinear terms can be neglected (RDT), that is when the micro-scale Froude number is much smaller
than 1, i.e. $Fr_{\eta}\equiv {u(\eta)}/{\mathcal{N} \eta} \ll 1$,
where $\mathcal{N}$ is the buoyancy (Brunt-V\"ais\"al\"a) frequency and $u(\eta)$ the
characteristic velocity fluctuation at the Kolmogorov length scale $\eta$,
the non-linear terms in Eqs~\ref{eqn:density} and \ref{eqn:lagstro}
can be neglected which leads to the
linearised Boussinesq equations:
\begin{eqnarray}
{D \over Dt} \mathbf{u} \simeq \frac{\partial}{\partial t}\mathbf{u}&=&-\frac{1}{\rho}\nabla p'+\frac{\rho'}{\rho} \mathbf{g}
\label{eqn:linear1}
\end{eqnarray}
where $\mathcal{N}^2 = g |d\rho/dx_3|\rho$.
The Fourier transforms $\tilde{\mathbf{u}}(\mathbf{k},t)$ of $\mathbf{u}(\mathbf{x},t)$
is used to solve Eq.~\ref{eqn:linear1}, so that the incompressibility
requirement is transformed into $\mathbf{k}\cdot\tilde{\mathbf{u}}(\mathbf{k},t)=0$ whilst the
pressure gradient is transformed into a vector parallel to
$\mathbf{k}$ in Fourier space. If $\mathbf{e}_3$ is the unit vector in
the direction of stratification, and $\mathbf{e}_1,\mathbf{e}_2$ are two unit vectors
normal to each other and to $\mathbf{e}_3$ ($\mathbf{g}=-g\mathbf{e}_3$),
the Craya-Herring frame (see Fig.~\ref{fig:craya}) is given by the unit vector
$\hat{\mathbf{k}}={\mathbf{k}}/{k}$ and $\mathbf{c}_1={\mathbf{e}_3\times
\mathbf{k}}/{|\mathbf{e}_3\times \mathbf{k}|}$, $\mathbf{c}_2={\mathbf{k}\times \mathbf{c}_1}/{|\mathbf{k} \times \mathbf{c}_1|}$.
\\[2ex]
In the Craya-Herring frame the Fourier transformed velocity field
$\tilde{\mathbf{u}}(\mathbf{k},t)$ lies in the plane defined by
$\mathbf{c}_1$ and $\mathbf{c}_2$ i.e.
\begin{equation}
\tilde{\mathbf{u}}(\mathbf{k},t)=\tilde{v}_1(\mathbf{k},t)\mathbf{c}_1+\tilde{v}_2(\mathbf{k},t)\mathbf{c}_2
\end{equation}
and is therefore decoupled from the pressure fluctuations which are
along $\mathbf{k}$. Incompressible solutions of Eq.~\ref{eqn:linear1}
in Fourier space and in the
Craya-Herring frame are (for the sake of simplicity, the initial potential is set to 0) \cite{Godeferd-Cambon94}:
\begin{eqnarray}
\tilde{v}_1(\mathbf{k},t) & = &
\tilde{v}_1(\mathbf{k},0)
\label{eqn:ini1}
\\
\tilde{v}_2(\mathbf{k},t)&=&\tilde{v}_2(\mathbf{k},0) \cos \sigma t
\label{eqn:ini2}
\end{eqnarray}
where $\sigma=\mathcal{N} \sin \theta$ and
$\theta=\theta(k)$ is the angle between $\mathbf{k}$ and vertical axis
$\mathbf{e}_3$. The initial conditions that we have to choose
are $\tilde{v}_1(\mathbf{k},0)$ and $\tilde{v}_2(\mathbf{k},0)$,
We emphasize again that the linearized Boussinesq equations are not valid if
$Fr_{\eta} \ll 1$ does not hold.

\subsection{Kinematic Simulation of stratified non decaying
turbulence}
The initial three dimensional turbulent velocity field used for the stably
stratified turbulence is taken from a homogeneous isotropic KS.
Using Fourier decomposition, the initial velocity
$\textbf{u}(\textbf{x},0)$ in spherical coordinates can be written as follows:
\begin{equation}
\textbf{u}(\textbf{x},0)=\int^{k_N}_{k_1}\int^{\pi}_0\int^{2\pi}_0\,\tilde{\textbf{u}}(\textbf{k},0)\,
k^2\sin\,\theta\,dk\,d\theta\,d\phi\,e^{i\textbf{k}\cdot
\textbf{x}} .
\label{eqn:inivel1}
\end{equation}
The initial KS velocity field is built by discretizing Eq.
\ref{eqn:inivel1}:
\begin{equation}
\textbf{u}(\textbf{x},0)=\sum^{N}_{n=1}\sum^{M}_{m=1}\sum^{J}_{j=1}\tilde{\textbf{u}}(\textbf{k}_{mnj},0)\,
k^2_{n}\sin\theta_m\Delta
k_n\Delta\theta_m\Delta\phi_j\,e^{i\textbf{k}_{mnj}\cdot\textbf{x}}
\end{equation}
where
$\textbf{k}_{mnj}=k_n(\sin\theta_m\cos\phi_j,\sin\theta_m\sin\phi_j,\cos\theta_m)$
(see \cite{Nicolleau-Vassilicos-2000}). For each pair $n,m$ we
randomly pick out one $\phi_j$, therefore
the notation $\phi_j$ should be replaced by $\phi_{nm}$ and the KS field reduces to
\begin{equation}
\mathbf{u}(\mathbf{x},0)=2\pi \mathcal{R}e
\left \{
\sum^N_{n=1}\sum^M_{m=1}\tilde{\mathbf{u}}
(\mathbf{k}_{mn},0)k^2_n \sin \theta_m \Delta
k_n\Delta\theta_me^{i\mathbf{k}_{mn}\cdot\mathbf{x}}
\right \}
\label{eqn:ksvel1}
\end{equation}
where $\mathcal{R}e$ stands for {\it real part} and
\begin{equation}
\tilde{\mathbf{u}}(\mathbf{k}_{mn},0)=\tilde{v}_1(\mathbf{k}_{mn},0)\mathbf{c}_1(\mathbf{k}_{mn},0)+\tilde{v}_2(\mathbf{k}_{mn},0)\mathbf{c}_2(\mathbf{k}_{mn})\label{eqn:ksv0}
\end{equation}
and
$\mathbf{k}_{mn}=k_n(\sin \theta_m \cos \phi_{nm},\sin \theta_m \sin \phi_{nm},\cos \theta_m)$.
%
\\[2ex]
%
Hence, there are $M$ wave vectors for a wavelength $k_n$.
%
The energy spectrum $E(k)$ is prescribed as follows:
\begin{equation}
\left \{
\begin {array}{ll}
E(k)=E_0 L(kL)^4 & \mbox{for $ k_1 < k \leq k_L$}
\\
E(k)=E_0L(kL)^{-5/3} & \mbox{for ${k_L} < k \le k_N $}
\\
E(k)=0 & \mbox{for $k_N < k $ and $k < k_1$}
\end{array}
\right .
\end{equation}
where $k_L=1/L$ and $L$ is the energy-containing length-scale.
The total kinetic energy of the turbulent fluctuation velocities scales with
$E_0$, and the Kolmogorov length-scale is represented by
$\eta=1/k_N$.
The wavenumbers are geometrically distributed i.e.
\begin{equation}
k_n=k_1\left(\frac{k_N}{k_1}\right)^{{n-1 \over N-1}} \mbox{ and
$k_1=\frac{1}{4L}$}
\end{equation}
%
\begin{equation}
\Delta k_n=\frac{k_n}{N-1} \ln \left(\frac{k_N}{k_1}\right) \Delta n .
\end{equation}
In order to capture correctly the effect of stratification, for each wavenumber,
$M$ wavevectors are defined such that
\begin{equation}
\theta_m=\frac{(m-1)}{M-1} \pi \, \mbox{ for $1 \le m \le M$}
\end{equation}
and
\begin{equation}
\Delta \theta_m=\frac{\pi}{M-1}\Delta m .
\end{equation}
Then, the angle in the horizontal plan $\phi_{mn}$ is chosen randomly in the range $[0, 2 \pi[$.
The velocity
field $\mathbf{u}(\mathbf{x},t)$ can be expressed at any time as
\begin{eqnarray}
\mathbf{u}(\mathbf{x},t)&=&
2\pi \mathcal{R}e
\left \{
\sum^N_{n-1}
\sum^M_{m=1}k^2_n \sin \theta_m \Delta
k_n\Delta\theta_me^{(i\mathbf{k}_{mn}\cdot\mathbf{x}+\omega_{mn}t)}
\right .
\\
& &
\left .
\times\left[\tilde{v}_1(\mathbf{k}_{mn},t)
\mathbf{c}_1(\mathbf{k}_{mn})+\tilde{v}_2(\mathbf{k}_{mn})\mathbf{c}_2(\mathbf{k}_{mn})\right]
\right \}
\label{eqn:ksvel}
\end{eqnarray}
$\tilde{v}_1(\mathbf{k}_{mn},t)$, $\tilde{v}_2(\mathbf{k}_{mn},t)$ 
obey Eqs \ref{eqn:ini1} and \ref{eqn:ini2} respectively.
A time-dependence $\omega_{mn}t$ has also been introduced in Eq.~\ref{eqn:ksvel}
in order to simulate time-decorrelation due to non-linearities
(see \cite{Nicolleau-Vassilicos-2000} for details and explanations on this point):
$\omega_{mn}=\lambda_{mn}\sqrt{k^3_nE(k_n)}$, where $\lambda_{mn}$
is a dimensionless unsteadiness parameter equal to 0.5
for all $m$ and $n$ in this paper.

\subsection{Particles with inertia}

In this paper we consider
particles of density $\rho_p$, moving in a fluid
of density $\rho$ and kinematic viscosity $\nu$.
The particles are heavy, that is their density $\rho_p$
is much larger than $\rho$ the density of the surrounding fluid.
Moreover, we assume
that the particles are rigid, have a spherical
shape characterised by a radius $a$ smaller than $\eta$ and are passive, that is they are transported by the flow without
affecting the flow. Furthermore, we assume that the particles are dilute enough not to interact with each other.

Under these assumptions, only the
drag and buoyancy forces are important as long as the particle
Reynolds number is much smaller than 1.
It can be shown \cite{Gatignol-1983,Maxey-Riley-1983} that the fluid force on such very small and heavy spherical particles is simply a linear Stokes drag force,
so that the position $\textbf{x}_p(t)$ of a particle and its velocity $\textbf{v}(t)$ at any instant are
related by Newton's second law in the simplified form
\begin{equation}
m_p \frac{d \textbf{v}}{dt}=6\pi a
\mu(\textbf{u}(\textbf{x}_p,t)-\textbf{v}(t))+m_p\textbf{g}
\label{maxou1}
\end{equation}
where $m_p$ is the particle's mass, $\mu$ is the
fluid's dynamic viscosity, $\textbf{g}$ is the gravitational
acceleration and $\textbf{u}(\textbf{x}_p,t)$ is the fluid velocity
at the position of the particle at time $t$.
Equation~\ref{maxou1} can be re-written as follows:
\begin{equation}
\frac{d}{dt}\textbf{v}=\frac{1}{\tau_p}[\textbf{u}(\textbf{x}_p,t)-\textbf{v}(t)]+\textbf{g}
\label{eqn:invel1}
\end{equation}
where $\tau_p$, the relaxation time, is
\begin{equation}
\tau_p = \frac{2\rho_p a^2}{9\rho \nu}
\label{deftaup}
\end{equation}
in terms of the kinematic viscosity $\nu = \mu / \rho$.
\\[2ex]
In order to calculate the particles' dispersion, we track the inertial particles in
time using
\begin{equation}
\dot{\mathbf{x}}_p(t)=\mathbf{v}(t)\label{eqn:tracking}
\end{equation}
We obtain the Lagrangian trajectories $\mathbf{x}_p(t)$ by integrating Eqs~\ref{eqn:tracking} and \ref{eqn:invel1}
using Eq.~\ref{eqn:ksvel} for the Eulerian flow velocity.
Each particle is released at a time $t_0$ from an initial position $x_0$
randomly chosen in each realization.
It is natural to define a drift velocity as follows
\begin{equation}
V_d=\tau_pg
\end{equation}
and the fall
velocity parameter (or drift parameter) $W$ as:
\begin{equation}
W=\frac{V_d}{u'}
\end{equation}
where $u'$ is the r.m.s. turbulence velocity.
%
%
In isotropic turbulence, $\tau_p$ needs to be
compared to $\tau_{\eta}$ and $L/u'$ (e. g. \cite{Maxey-1987,Wang-Maxey-1993,Fung-1998,Yang-Lei-1998}).
Note that here $Fr = u' /NL \ll 1$ because $Fr_{\eta} \ll 1$ (in fact $ Fr \ll Fr_{\eta}$ in the high Reynolds number limit).
Our assumption $a \ll \eta$ imposes
\begin{equation}
a^2 \ll \eta^2
\end{equation}
that is
\begin{equation}
{a^2 \over \nu} {\rho_p \over \rho} < {\eta \over u_{\eta}} {\eta u_{\eta} \over \nu} {\rho_p \over \rho}
\end{equation}
By definition of the Kolmogorov scale $\eta u_{\eta} / \nu=1$,
that is
\begin{equation}
\tau_p \ll \tau_{\eta} {\rho_p \over \rho}
\end{equation}
As under our assumption of heavy particle $\rho \ll \rho_p$, we chose to limit the study of the paper to
\begin{equation}
\tau_p \le \tau_{\eta}
\label{eqn:taup}
\end{equation}
\\[2ex]
The Boussinesq approximation we are basing our KS on,
requires that the vertical thickness of a layer of stratified fluid is
small enough for the mean density $\rho$ and the mean density
gradient ${d\rho}/{dx_3}$ to be effectively independent of $x_3$
within that layer, and the thickness of this layer can be estimated
as much smaller than $H \equiv \rho/|d\rho/dx_3|={g}/{\mathcal{N}^2}$. The KS
turbulence model we consider here can therefore only make sense if the
integral scale of the turbulence is much smaller than $H$, i.e.
$H \gg L$. This leads to
\begin{equation}
\frac{1}{\mathcal{N}} \gg \sqrt{\frac{L}{g}} .
\end{equation}
The low
Froude number condition on which we have based the linearisation of the Boussinesq equation, $Fr_{\eta}={u_{\eta}}/{\eta \mathcal{N}}=(1/\tau_{\eta})\mathcal{N} \ll 1$,
implies ${1}/\mathcal{N} < \tau_{\eta}$. Adding
to these conditions our high Reynolds number limit
(${L}/{\eta} \gg 1$), we have a set of time scales ordered as
follows:
\begin{equation}
\sqrt{\frac{\eta}{g}}
<
\sqrt{\frac{L}{g}}
<
\frac{1}{\mathcal{N}}
<
\tau_{\eta}
<
\frac{L}{u'}
\label{cond1} .
\end{equation}
Inertial particles are characterized by their relaxation time $\tau_p$ and
different particle behaviours may be observed
depending on the relation between $\tau_p$ and the different characteristic times in Eq.~\ref{cond1}. This leads to the different relaxation time
regimes we are investigating in this paper. We study the behaviour of inertial particles by setting
$\tau_p$ to lie within these different relaxation time regimes and
changing the drift parameter $W=\tau_p g/u'$ to $W>1$ or $W<1$.
\\[2ex]
The general parameters for the different KS runs of stratified turbulence are
presented in Table~\ref{tableKS}.
\begin{table}[h]
\caption{
\label{tableKS}
KS's parameters.
}
\begin{tabular}{rrcrcc}
\hline\noalign{\smallskip}
$L/\eta$ & $\eta$ \hspace*{0.5cm} & $L/u'$ & $\tau_{\eta}$ \hspace*{0.5cm} & $\sqrt{\eta / g}$ & $\mathcal{N}$
\\
\noalign{\smallskip}\hline\noalign{\smallskip}
100 & $10^{-2}$ & 1 & $4.6 \times 10^{-2}$ & $3.16 \times 10^{-5}$ & 500, 1250, 2000, 2500, 3000
\\
4000 & $2.5 \times 10^{-4}$ & 1 & $3.97 \times 10^{-3}$ & $1.58 \times 10^{-5}$ & 500, 1250, 2000, 2500, 3000
\\
\noalign{\smallskip}\hline
\end{tabular}
\end{table}
There are two sets of Eulerian parameters; for each, $\mathcal{N}$ and $\tau_p$ were varied.
Our choice of
parameters always satisfy condition~(\ref{cond1}).
Not all cases are shown in this paper, there would have been too many. We have chosen to show only representative runs for each cases.

\subsection{Simulations}

We performed simulations by releasing particles characterized by an inertial characteristic time
$\tau_p$ into the strongly stratified turbulence.
The initial
position of a particle $(x_0,y_0,z_0)$ is chosen randomly in each realization. The
time step $\Delta t$ is chosen such that $\Delta t$ is smaller than $\tau_p$ and
$\sqrt{{\eta}/{g}}$. The unsteadiness frequency parameter
$\lambda$ is set to $0.5$. The equation of motion is
integrated for 2000 realizations of the flow field. By different
realizations we mean different trajectories in different velocity
field realizations. The initial condition for the equation of
motion of the particle is
\begin{equation}
\textbf{v}(t=0)=\textbf{u}_3(\textbf{x}_p(t=0),0)- \mathbf{V}_d .
\end{equation}
However, in all figures, the starting time zero is
$t_0=(10-\delta)2\pi/\mathcal{N}$ where $\delta$ is a different random number
between -1 and 1 for different trajectories in order to avoid an
initial in-phase oscillation of particles together and allow what
may be a more realistic particle release.
The relative time used in the figures is $\tau = t-t_0$.
In this way, all our results
correspond to times after which stratification has had time to be
established in our KS velocity field (see \cite{Nicolleau-Vassilicos-2000}).
The equation of motion was integrated using a 4th order
Runge-Kutta method.
\\[2ex]
The relaxation time $\tau_p$ is independent of the vertical position because the
changes of mean density with altitude and depth in a strongly
stratified Boussinesq turbulence are negligible as
$\partial\rho/\partial x_3$ remains small under the Boussinesq
approximation.

\subsection{Fluid particle KS simulations}

For the sake of comparison it may be worth summarising the main results obtained for one-particle diffusion in isotropic and stratified Kinematic Simulation.
See e.g. \cite{Nicolleau-Vassilicos-2000} for a more complete discussion of fluid particle diffusion in
KS stratified flows. In isotropic or stratified turbulence, if there is a mean flow $\mathbf{V}_d$,
the mean departure from the initial position for a fluid particle is given by:
\begin{equation}
\langle\mathbf{x}-\mathbf{x}_0\rangle = \mathbf{V}_d (t-t_0)
\end{equation}
The rms of the departure at small time is given by the ballistic regime.
\begin{equation}
<(x_3(t) - x_3(t_0))^2> \sim {u'}^2 (t-t_0)^2
\end{equation}
for $t-t_0 \ll {L /u'}$ in the case of an isotropic turbulence.
That ballistic regime has a different duration in the vertical direction for
a stratified flow. In this case, it is valid for $t-t_0 \ll {2 \pi / \mathcal{N}}$.
For large times, the fluid particle in isotropic turbulence follows a random walk regime:
\begin{equation}
<(x_i(t) - x_i(t_0))^2> \sim {u'} L (t - t_0)
\mbox{ \hspace*{1cm} for }\, t-t_0 > {L \over u'}
\end{equation}
whereas its diffusion in the vertical direction is capped in stratified turbulence:
\begin{equation}
\langle({x_3}-{x_3}(0))^2\rangle = 1.5 {{u'}^2 \over {\mathcal{N}}^2} + \mathrm{oscillations} \mbox{ \hspace*{0.55cm} for }\, t - t_0 > {2 \pi \over N}
\label{fp3stra}
\end{equation}
It may appear as a paradox that a synthetic flow without the anisotropic Eulerian structuration found in DNS - often refered to as `pancake' structure' - can reproduce accurately
the anisotropic Lagrangian dispersion.
The apparent paradox comes from the misleading comparison of the Lagrangian
`two-time' correlation with the single-time two-point Eulerian correlation. The
linear operator coming from the RDT assumption (e.g Eq. \ref{eqn:ini2}) gives rise to the important phase terms $e^{\pm i\sigma(k)t}$. Time dependency
can cancel out for single-time two-point Eulerian velocity auto-correlations, if started
from isotropic initial data. With the phase term multiplied by its complex conjugate,
the necessary oscillations leading to Lagrangian anisotropy cannot be created by the purely linear solution. Whereas, an
anisotropic evolution is possible from two-time Eulerian velocity auto-correlations
by multiplying $e^{\pm i\sigma(k)t}$ by its complex conjugate at another time $t_0$. This is the key to the use of the simplified Corrsin's hypothesis
\cite{Cambon-al-2004,Nicolleau-Vassilicos-2000,Nicolleau-Yu-2007,Nicolleau-al-2008}.
\\[2ex]
Let us now consider the effect of inertia on these different regimes. We start with inertial particles with small responsive times
in section \ref{sectoupetit} ($\tau_p<\sqrt{{\eta}/{g}}$) and increase $\tau_p$ up to the limit of our model validity ($\sqrt{{L}/{g}}\leq\tau_p\leq\tau_{\eta}$)
in section~\ref{sec:sec2}.

\section{First regime: very small response time, $\tau_p<\sqrt{{\eta}/{g}}$}
\label{sectoupetit}

In this section we consider cases where
\begin{equation}
\tau_p < \sqrt{\frac{\eta}{g}}
\label{condf1}
\end{equation}
which with (\ref{cond1}) means $\tau_p < \sqrt{\eta/g} < \sqrt{L/g} < 1/\mathcal{N}<\tau_{\eta}<L/u'$.
$\sqrt{{\eta}/{g}}$ can be thought of as the characteristic time to fall through an eddy of size $\eta$ under the effect of gravity. It seems then natural to define a small scale gravity effect Stokes number $St_{g \eta}$ and a large scale gravity effect Stokes number $St_{gL}$ as follows
\begin{equation}
St_{g \eta} = \tau_p \sqrt{\frac{g}{\eta}}, \,\,
St_{gL} = \tau_p \sqrt{{g \over L}}
.
\end{equation}
The cases under consideration here correspond to $St_{g \eta} < 1$ and of course also $St_{gL}<1$
and the set of conditions~\ref{cond1}. These different cases are detailed in Table~\ref{tableI}.
\begin{table}[h]
\caption{
\label{tableI}%
Cases used in section~\ref{sectoupetit}.}
%
\begin{tabular}{p{5cm}p{5cm}}
\hline\noalign{\smallskip}
\begin{tabular}{crrrr}
case & $\tau_p$ & $W$ & $ \mathcal{N} $ & $St_{g \eta}$
\\
\noalign{\smallskip}\hline\noalign{\smallskip}
A\ref{sectoupetit} & $5 \times 10^{-7}$ & $ 0.5 $ & 500 & 0.015
\\
B\ref{sectoupetit} & $3 \times 10^{-6}$ & $ 3.0 $ & 500 & 0.095
\\
C\ref{sectoupetit} & $ 10^{-5}$ & $ 10.0 $ & 500 &0.316
\\
D\ref{sectoupetit} & $5 \times 10^{-7}$ & $ 0.5 $ & $ 1250 $ & 0.015
\\
E\ref{sectoupetit} & $3 \times 10^{-6}$ & $ 3.0 $ & $ 1250 $ & 0.095
\end{tabular}
&
\begin{tabular}{crrrr}
case & $\tau_p$ & $W$ & $ \mathcal{N} $ & $St_{g \eta}$
\\
\noalign{\smallskip}\hline\noalign{\smallskip}
F\ref{sectoupetit} & $ 10^{-5}$ & $ 10.0 $ & $ 1250 $ & 0.316
\\
G\ref{sectoupetit} & $5 \times 10^{-7}$ & $ 0.5 $ & $ 2500 $ & 0.015
\\
H\ref{sectoupetit} & $3 \times 10^{-6}$ & $ 3.0 $ & $ 2500 $ & 0.095
\\
I\ref{sectoupetit} & $ 10^{-5}$ & $ 10.0 $ & $ 2500 $ & 0.316
\\
\\
\end{tabular}
\\
\noalign{\smallskip}\hline
\end{tabular}
%
\end{table}

\subsection{Mean displacement}

\begin{figure}[h]
 \includegraphics[width=7cm]{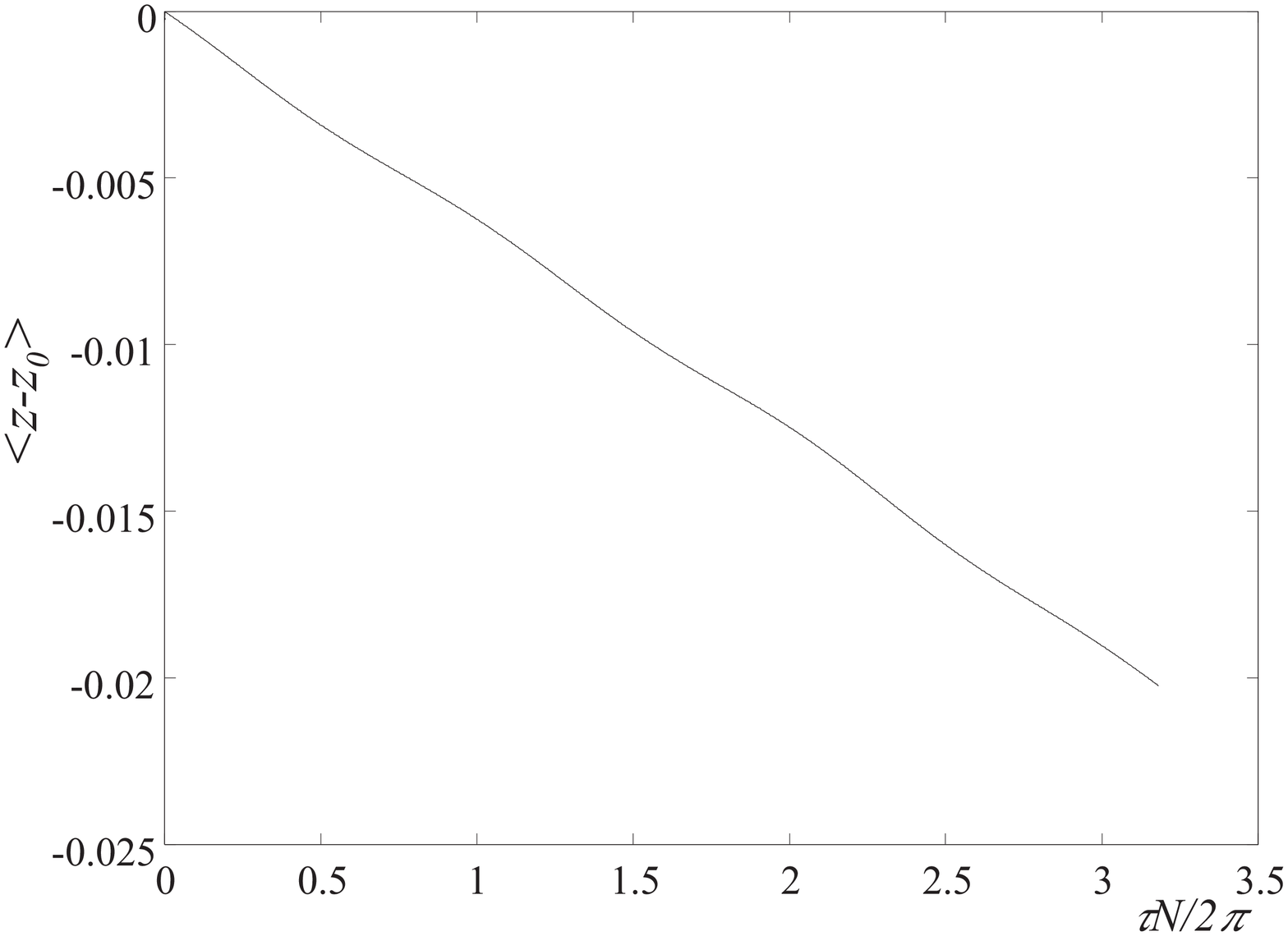}
\caption{%
\label{fig:fig101}
$\langle z \rangle \simeq \langle z_0 \rangle - V_d \, \tau$ as a function of
$\tau \mathcal{N} / 2 \pi$
when
$\tau_p$ is less than $\sqrt{\eta/g}$ case A\ref{sectoupetit} in table~\ref{tableI}.}
\end{figure}

Fig.~\ref{fig:fig101} shows the vertical mean displacement $\langle z-z_0 \rangle$ as a function of time for case A\ref{sectoupetit}
in table~\ref{tableI}.
It shows clearly that $\langle z \rangle$ decreases linearly with time.
That is
even if $\tau_p$ is extremely small compared to
$\tau_{\eta}$, and in fact smaller than $\sqrt{\eta/g}$ in this
section, particles with inertia fall down linearly with a gradient
$V_d = \tau_p \ g$,
\begin{equation}
\langle z \rangle \simeq \langle z_0 \rangle - V_d \, \tau
.
\label{av1eqz}
\end{equation}
This is observed for $W= V_d/ u'$ as small as 0.5.
This result may not be too surprising as although $\tau_p$ is very small,
the inertial particles are still much
heavier than the fluid elements.
There is no effect of stratification as without stratification in the sole presence of gravity the particle will also move according to
(\ref{av1eqz}) in the vertical direction.
Thus, when $\tau_p < \sqrt{\eta/g}$ the inertial particle motion remains dominated by gravity provided that $W \ge 0.5$.
The same result has been obtained for the other cases in Table~\ref{tableI} (not shown here).

%

\subsection{Relative departure variance}
\label{IIIA2}

\begin{figure}[h]
\includegraphics[width=7cm]{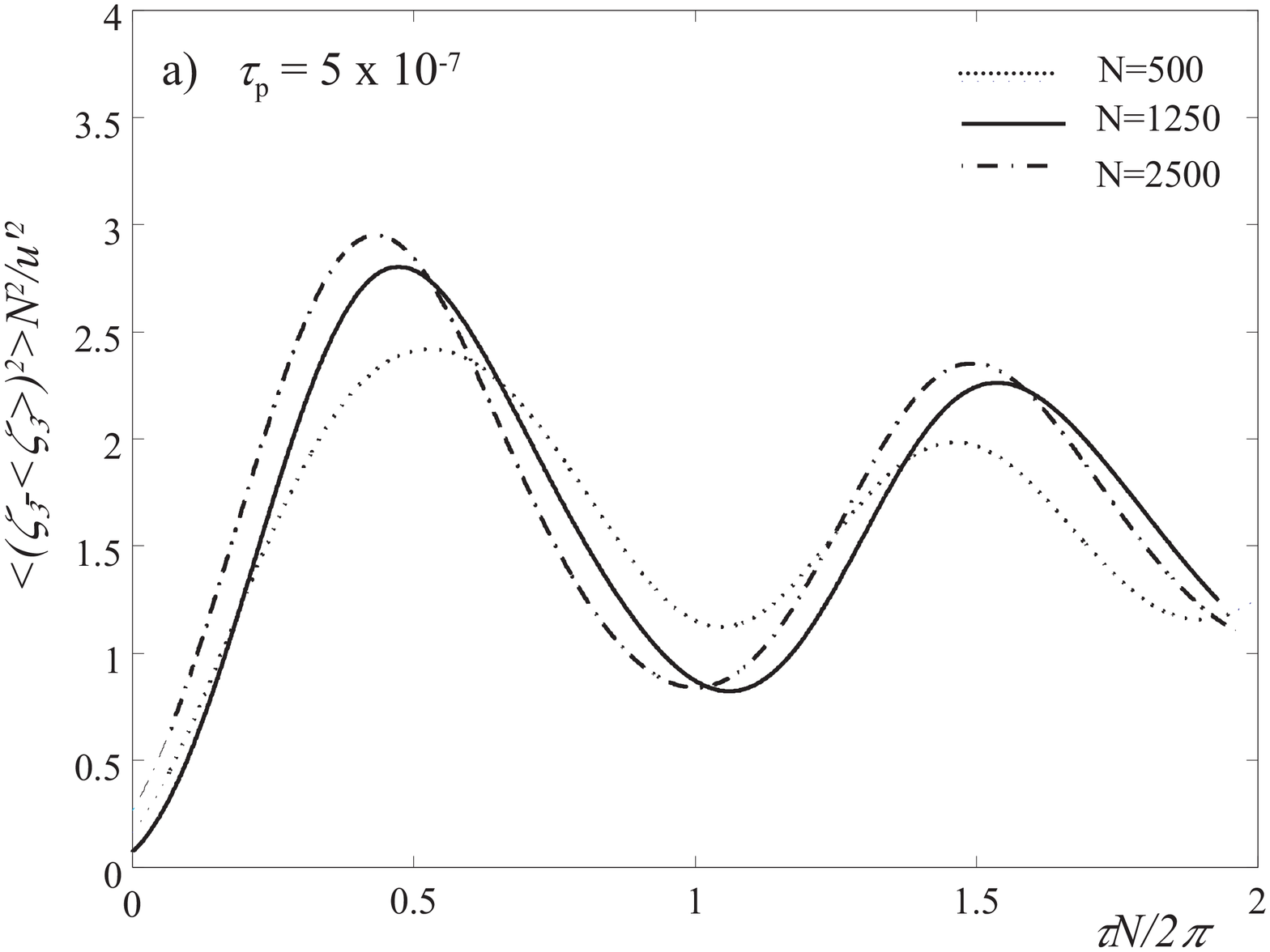}
\includegraphics[width=7cm]{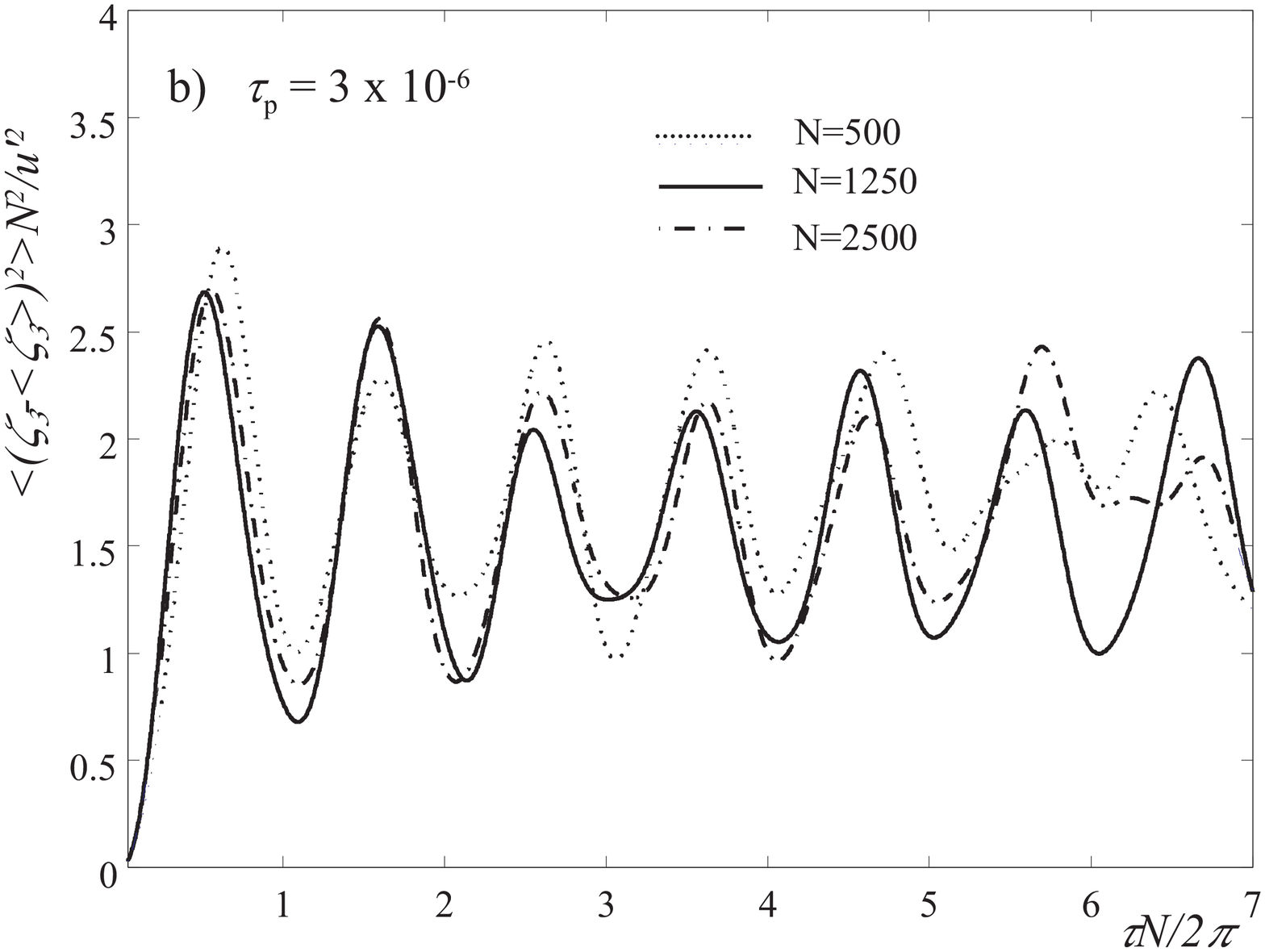}
\caption{%
\label{fig:varcollap1}%
$\langle (\zeta_3 -\langle \zeta_3 \rangle)^2 \rangle \mathcal{N}^2/u'$ as a function of $\tau \mathcal{N} / 2 \pi$
with different buoyancy frequencies when
$\tau_p < \sqrt{{\eta}/{g}}$ and
a) $V_d/{u'}<1$;
b) $V_d/{u'}>1$.
}
\end{figure}

\begin{figure}[h]
 \includegraphics[width=6cm]{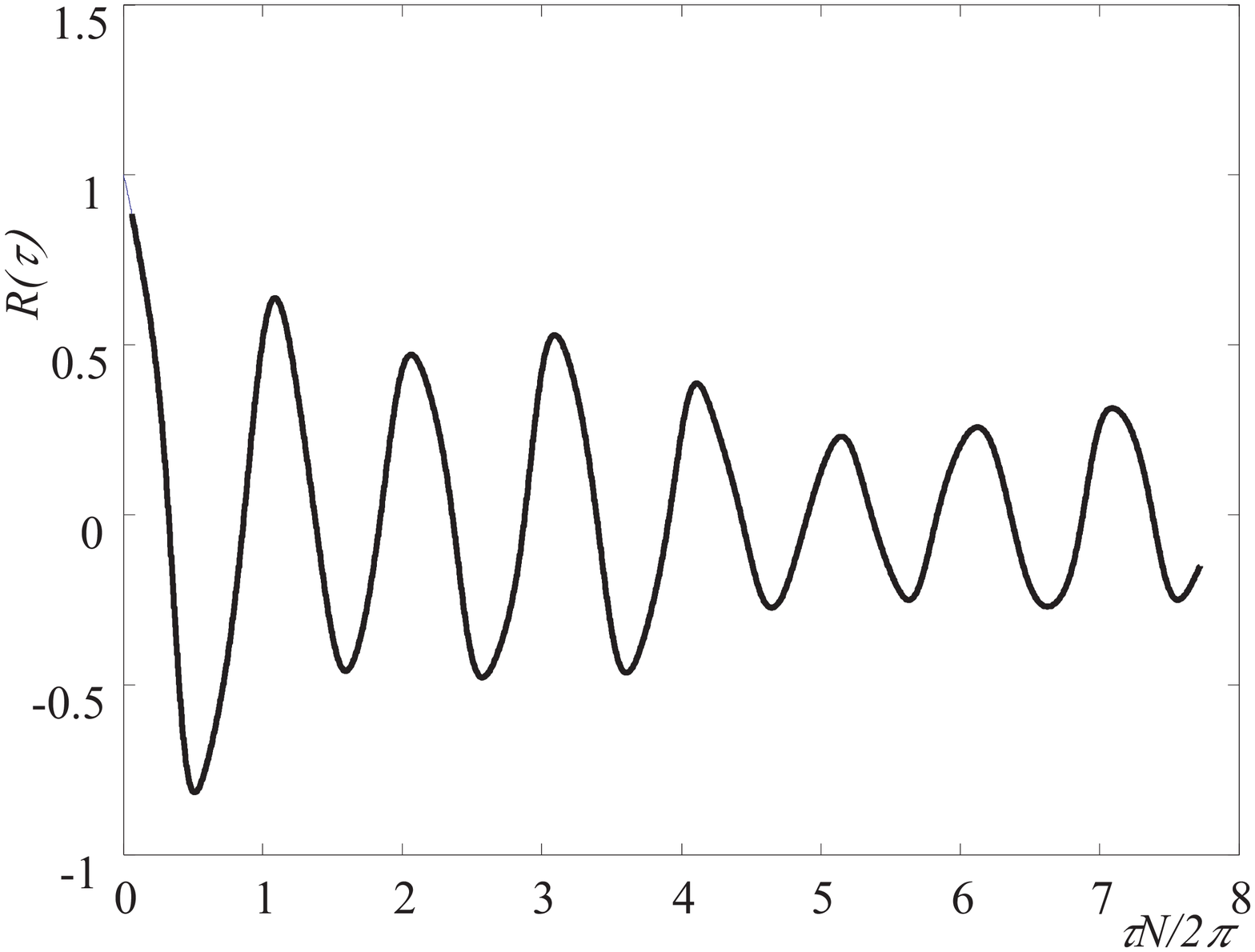}
\caption{\small\label{fig:fig104}
$R(\tau)$ as a function of $\tau \mathcal{N} /2 \pi$
when $\tau_p < \sqrt{{\eta}/{g}}$ and
$W > 1$ , case B\ref{sectoupetit} in table~\ref{tableI}.}
\end{figure}

We define the vertical relative displacement $\zeta_3 = z-z_0$.
The variance of the vertical inertial particle position $\langle(\zeta_3 -\langle
\zeta_3 \rangle)^2 \rangle$ is shown in
Fig.~\ref{fig:varcollap1}.
It oscillates about a constant. This results holds in fact for $0.5 \le W \le 10$.
Furthermore, in both cases $W < 1$ and $W >1 $,
the variance of the vertical inertial particle position also collapses when normalised by $u'$ and $\mathcal{N}$ following the law
(\ref{fp3stra})
observed for fluid particle in
stratified turbulence
%
We find that
\begin{equation}
\langle(\zeta_3-\langle \zeta_3 \rangle)^2 \rangle \simeq 1.5 \frac{u'^2}{\mathcal{N}^2} + \mbox{ oscillations}
\label{z2osc}
\end{equation}
which is identical, to the behaviour of fluid particles reported in \cite{Nicolleau-Vassilicos-2000},
although the collapse is not as
good as that observed for fluid particles.
By contrast to the mean displacement which follows that of a heavy particle in an isotropic KS, the variance of the departure follows the behaviour of a fluid particle
in a stratified KS.
\\[2ex]
As discussed in \cite{Nicolleau-Yu-2007} the vertical capping of the particle dispersion is governed by the oscillations of the
velocity autocorrelation function.
The {normalised}
autocorrelation function $R(\tau)$ is defined as:
\begin{equation}
R(\tau)=\frac{\langle (v_{3}(t_0)- \langle v_{3}(t_0) \rangle) (v_{3}(t_0+\tau)-
\langle v_{3}(t_0+\tau) \rangle ) \rangle}{\langle (v_{3}(t_0)- \langle v_{3}(t_0))^2 \rangle}
\end{equation}
It is shown in
Fig.~\ref{fig:fig104}.
The particle diffusivity is defined as
\begin{equation}
\frac{d}{dt}\langle(\zeta_3 -\langle \zeta_3 \rangle)^2 \rangle
\end{equation}
and Taylor's (1921) relation states that it equals the integral of the velocity time autocorrelation function, i.e.
\begin{equation}
\frac{d}{dt} \langle(\zeta_3 -\langle \zeta_3 \rangle)^2\rangle=
{\langle (v_{3}(t_0)- \langle v_{3}(t_0) \rangle)^2 \rangle}
\int^t_0 R(\tau)d\tau
\label{taylorlise}
\end{equation}
The autocorrelation function is shown in Fig.~\ref{fig:fig104} to oscillate
around zero with the oscillation's amplitude decreasing with time. It is very similar to that found for a fluid particle in
\cite{Cambon-al-2004}.
Thus, according to Taylor's relation,
the vertical inertial particle diffusivity can also oscillate around zero.

Inertial particles with $\tau_P < \sqrt{\eta/g}$ disperse similarly to fluid particles whether $W$ is larger or smaller than 1 which
does not mean that inertial particles are fluid elements but that
they vertically diffuse like fluid elements if we
remove the falling effect.
%

\section{Second regime: in intermediate small inertial response times, $\sqrt{{\eta}/{g}}\leq\tau_p\leq\sqrt{{L}/{g}}$
\label{sec:43}}

\begin{table}[h]
\caption{\label{tableIII}%
Cases studied in section~\ref{sec:43}, in all cases $L=1$, $u'=1$.}
\begin{tabular}{l|l|r|r|r|r|r}
\hline\noalign{\smallskip}
case & $\tau_p$ & $W$ & $ \mathcal{N} $
& ${L}/{V_d}$ & ${2\pi}/{\mathcal{N}}$ & $Fr_d$
\\
\noalign{\smallskip}\hline\noalign{\smallskip}
A\ref{sec:43}
&$1.7 \times 10^{-5}$& 17&2000&
0.588&$3.14 \times 10^{-3}$&0.053
\\
B\ref{sec:43} &$1.7 \times 10^{-5}$& 17&2500&
0.588&$2.51 \times 10^{-3}$& 0.043
\\
C\ref{sec:43} &$1.7 \times 10^{-5}$& 17&3000&
0.588&$2.09 \times 10^{-3}$& 0.036
\\
D\ref{sec:43} & $5 \times 10^{-4}$& 500&500&
0.002&$1.26 \times 10^{-2}$& 6.28
\\
E\ref{sec:43} & $5 \times 10^{-4}$ & 500&1250&
0.002&$5.03 \times 10^{-3}$ & 2.51
\\
F\ref{sec:43} & $5 \times 10^{-4}$ & 500&2500&
0.002&$2.51 \times 10^{-3}$ & 1.26
\\
G\ref{sec:43} &$9 \times 10^{-4}$ & 900& $ 500 $ & 0.001 & $1.26 \times 10^{-2}$ & 11.46
\\
H\ref{sec:43} &$9 \times 10^{-4}$ & 900& $ 1250 $ & 0.001 & $5.03 \times 10^{-3}$ & 4.57
\\
I\ref{sec:43} &$9 \times 10^{-4}$ & 900& $ 2500 $ & 0.001 & $2.51 \times 10^{-3}$ & 2.28

\\
\noalign{\smallskip}\hline
\end{tabular}
\end{table}

In this section, $\tau_p$ is chosen in the intermediate time regime
$\sqrt{\eta/g}<\tau_p<\sqrt{L/g}$,
that is $St_{g \eta} > 1$ but $St_{g L} < 1$. Note that when $\tau_p > \sqrt{\eta/g}$, $W$ cannot be smaller than 1, so that $V_d$ is always larger than $u'$
in this regime.
The time that it takes for $v_3(t)-\langle v_3(t)\rangle$ to decorrelate
should be of the order of ${L}/{V_d}$ because $V_d \gg u'$ and particles
fall with an average fall velocity $V_d$ through eddies of
all sizes, the largest being $L$.
We therefore distinguish between two potential cases:
\begin{itemize}
\item[i)]
strong stratification
$1 /\mathcal{N}< {L}/{V_d}$ and
\item[ii)]
weak stratification
$1/{\mathcal{N}} >{L}/{V_d}$.
\end{itemize}
Then, we can introduce a new characteristic Froude number as follows:
\begin{equation}
Fr_{d} = \frac{V_d}{\mathcal{N} L} = \frac{u'}{L\mathcal{N}} \frac{V_d}{u'} = Fr \times W > Fr
.
\end{equation}
%
The flow
parameters used in the various simulations used to infer the conclusions reported here are shown in table~\ref{tableIII}.

\subsection{Stratification dominated sub-regime, $Fr_d < 1$}
\label{sdd}
\begin{figure}[h]
 \includegraphics[width=7cm]{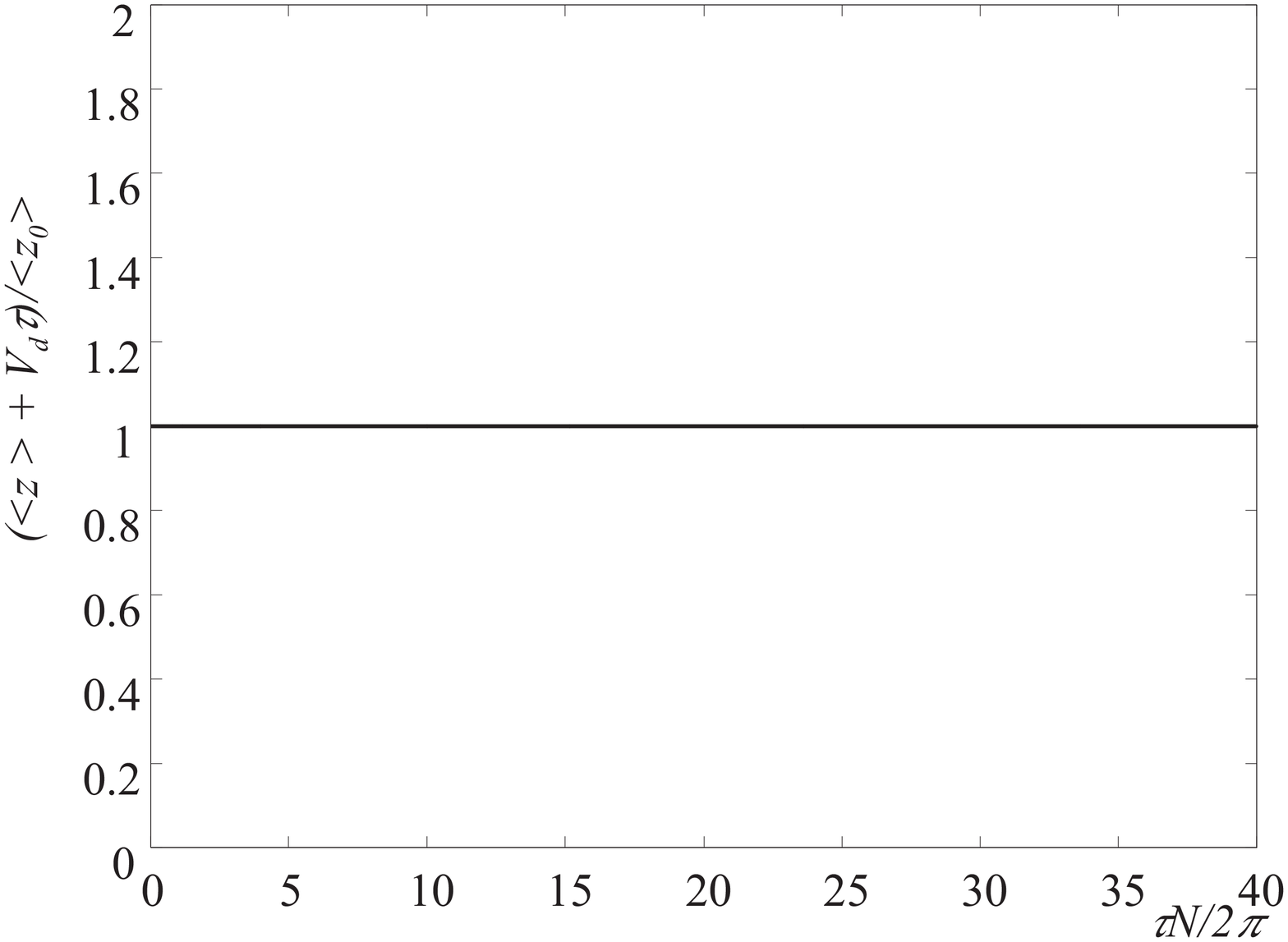}
\caption{\label{fig:grad2}%
$(\langle z \rangle + V_d \, \tau)/{\langle z_0 \rangle}$ when
 $\sqrt{{\eta}/{g}}\leq\tau_p\leq\sqrt{{L}/{g}}$, ${V_d}/{u'}>1$
for
$Fr_d < 1$ (case B\ref{sec:43} in table~\ref{tableIII}).
}
\end{figure}

\begin{figure}[h]
 \includegraphics[width=8cm]{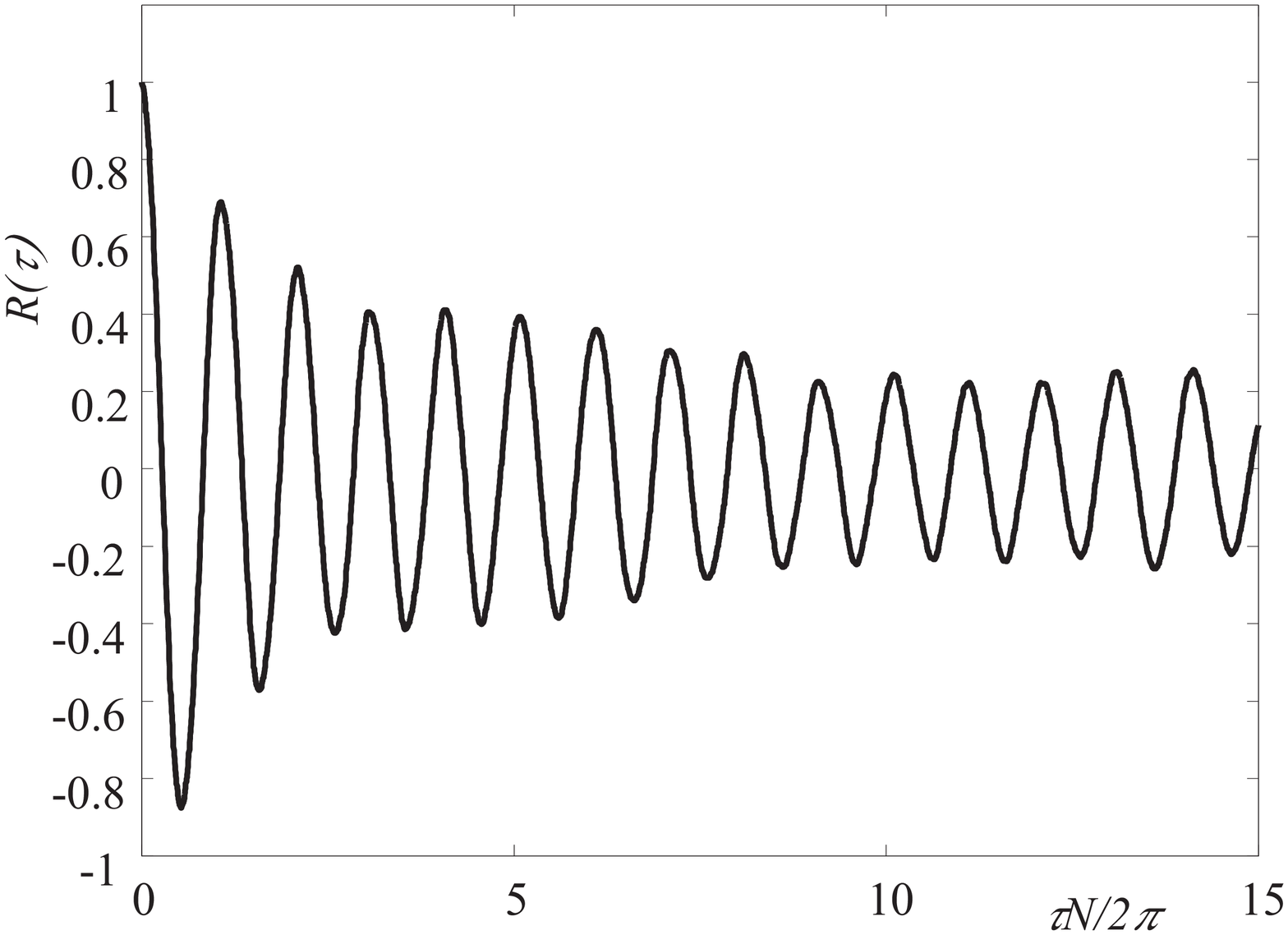}
\caption{\label{fig:cor1-1}%
$R(\tau)$
when $\sqrt{{\eta}/{g}}<\tau_p<\sqrt{{L}/{g}}$ for
$Fr_d < 1$. Case B\ref{sec:43} in table~\ref{tableIII}.
}
\end{figure}

\begin{figure}[h]
 \includegraphics[width=7cm]{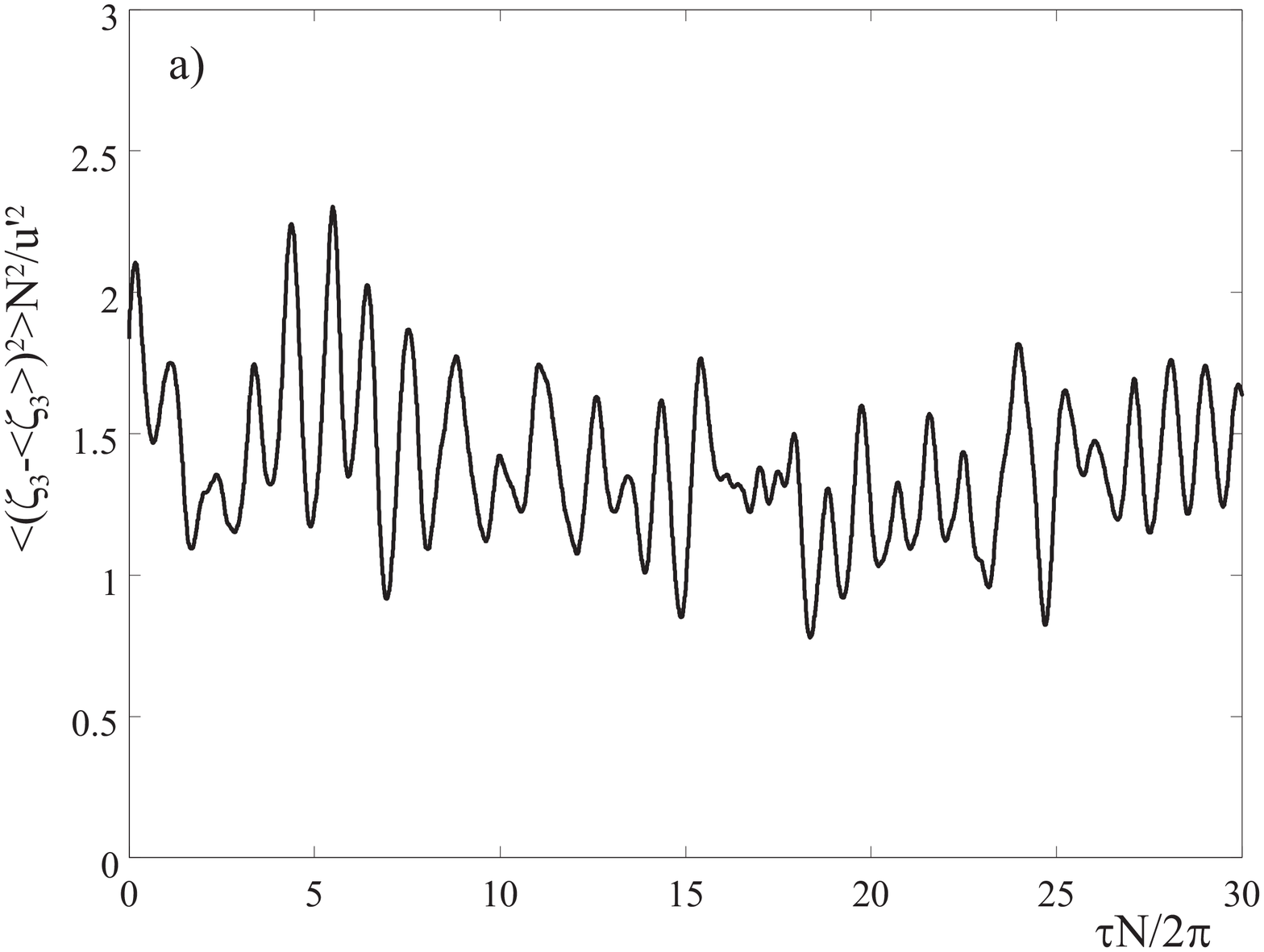}
 \includegraphics[width=7cm]{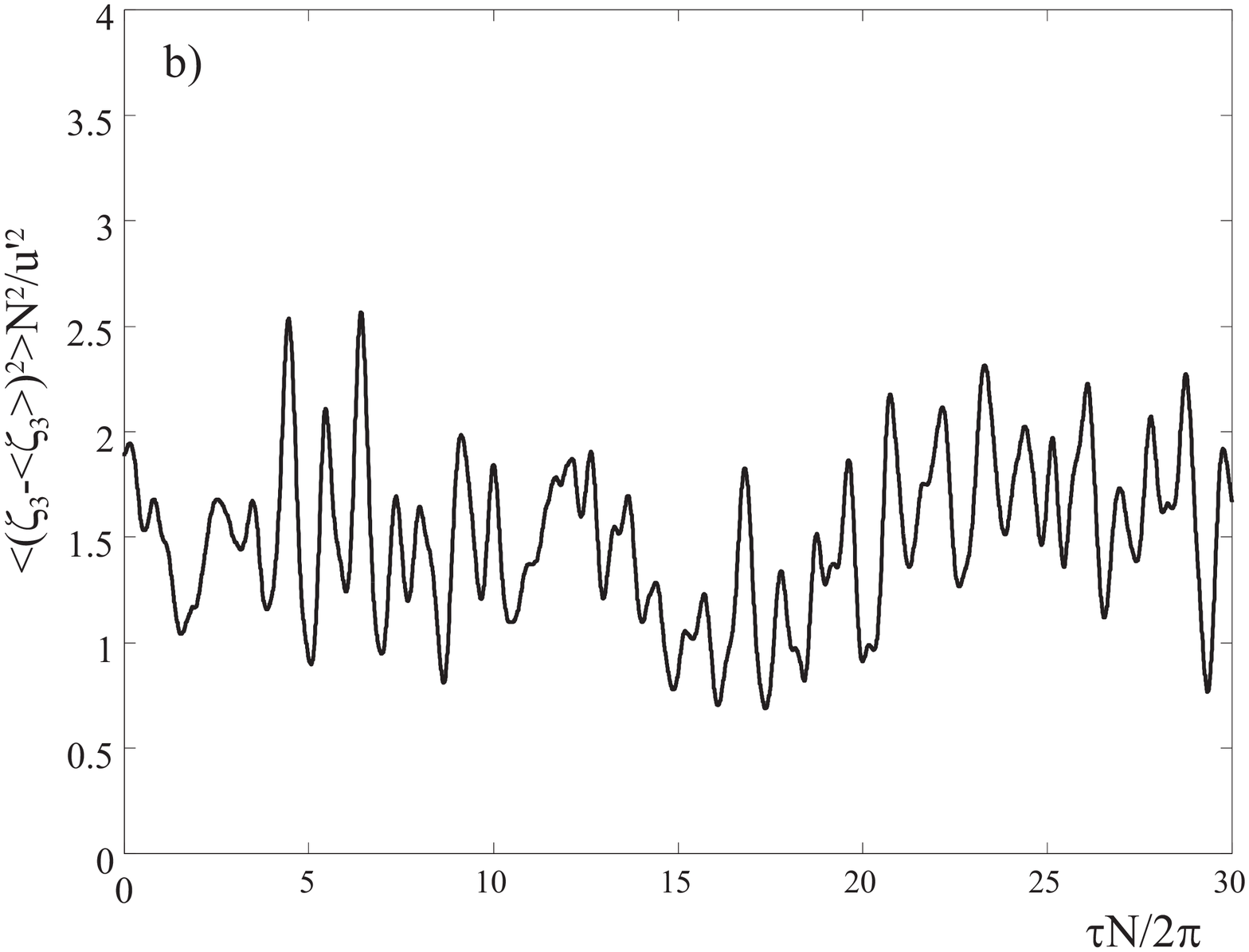}
\caption{%
\label{fig:fig107-1}
${\langle(z-\langle z\rangle)^2\rangle} \mathcal{N}^2/{u'^2}$ when
$\sqrt{{\eta}/{g}}<\tau_p<\sqrt{{L}/{g}}$ for
$Fr_d < 1$
a) case A\ref{sec:43}, b) case C\ref{sec:43} in Table~\ref{tableIII}.}
%
\end{figure}

%
In Fig.~\ref{fig:grad2} we see that inertial
particles still fall down with velocity $V_d$ when ${2\pi}/\mathcal{N}< {L}/{V_d}$
(results are identical for cases A\ref{sec:43}, B\ref{sec:43} and C\ref{sec:43}).
%
\\[2ex]
The centered variance of the inertial particle vertical relative position
$\langle (\zeta_3 -\langle \zeta_3\rangle)^2 \rangle \mathcal{N}^2/u'^2$
can again be deduced
from the autocorrelation function
of the particle vertical velocity using Taylor's relation.
Fig.~\ref{fig:cor1-1} shows
$R(\tau)=\langle (v_{3}(t_0)-\langle v_{3}(t_0) \rangle)(v_{3}(t_0+\tau)- \langle v_{3}(t_0+\tau) \rangle)\rangle/{\langle (v_{3}(t_0)- \langle v_{3}(t_0))^2 \rangle}$ as a function of $\tau \mathcal{N}/ 2 \pi$
(cases A\ref{sec:43}, B\ref{sec:43}, C\ref{sec:43} show identical results).
%
It is clear from Fig.~\ref{fig:cor1-1},
that this autocorrelation is dominated by gravity-wave oscillations and
Taylor's relation yields
\[
\frac{d}{dt}
\langle (\zeta_3 -\langle \zeta_3 \rangle)^2 \rangle =
{\langle (v_{3}(t_0)- \langle v_{3}(t_0))^2 \rangle}
\int^{\tau}_0 R(\tau')d\tau'\simeq0.
\]
Hence, we can conclude that the vertical diffusivity is 0 when $Fr_d < 1$ and as a consequence the variance of the vertical separation is bounded.
\\[2ex]
This variance of the inertial particle vertical relative position
can be calculated directly and is shown
in Fig.~\ref{fig:fig107-1}. It is found that $\langle(\zeta_3 -\langle \zeta_3 \rangle)^2\rangle =1.5 u'^2/\mathcal{N}^2 + oscillation$
as for $\tau_p < \sqrt{{\eta}/{g}}$ in section~\ref{IIIA2} (see Eq.\ref{IIIA2}).
It is worth noting that the oscillations are more irregular than for the cases $\tau_p < \sqrt{{\eta}/{g}}$ in section~\ref{IIIA2}.


\subsection{Gravity-dominated regime, $Fr_d >1$}
\label{weak4}

We repeat the calculations of the previous section but this time for $Fr_d >1$, that is
$2 \pi / \mathcal{N} > L / V_d$. Note that we are still in the case $Fr_{\eta} \ll 1$ as in the entire paper.
The stratification effect is weaker than in section~\ref{sdd} but remains very strong.
\\[2ex]
\begin{figure}[h]
 \includegraphics[width=7cm]{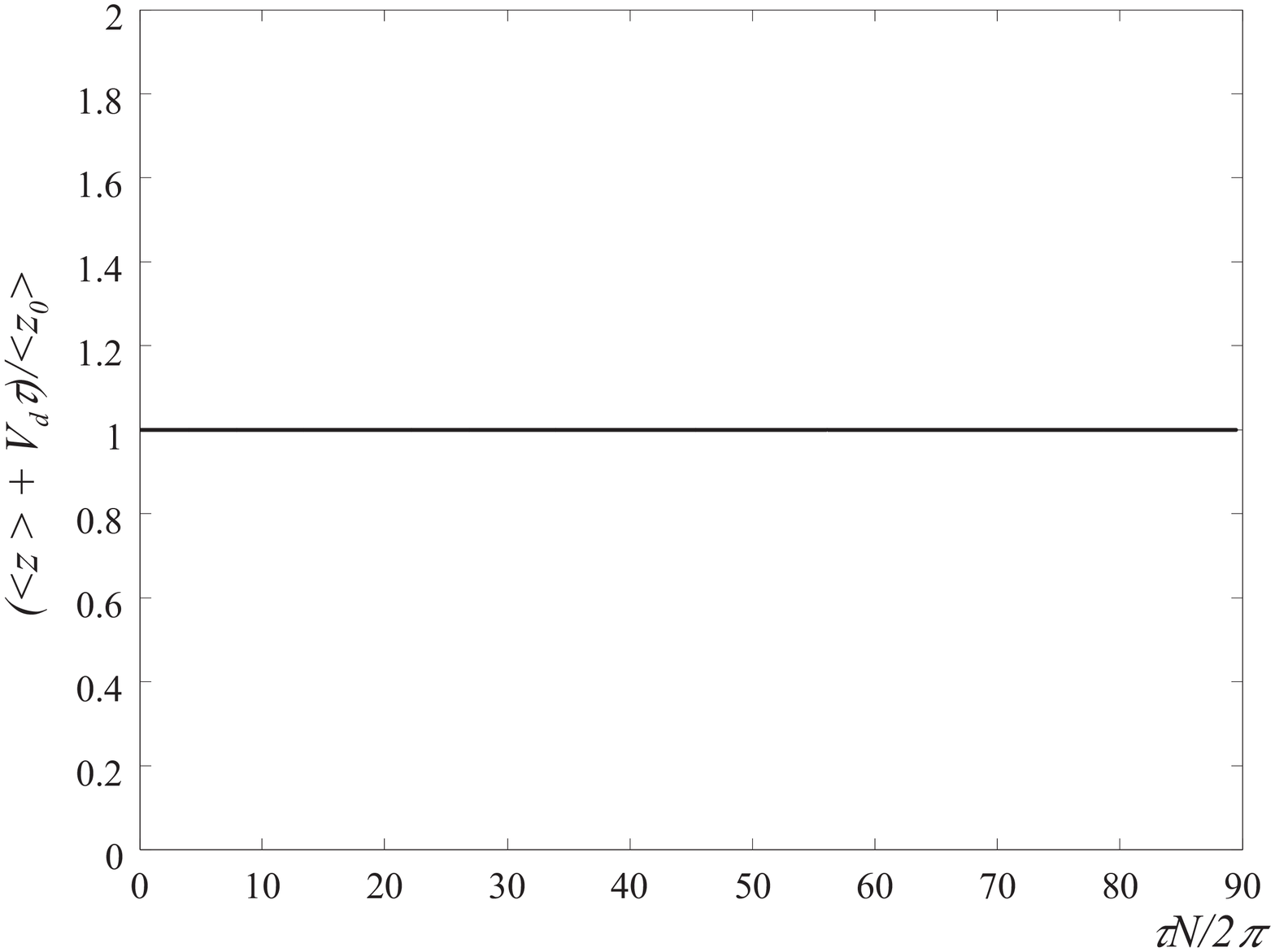}
\caption{\label{fig:grad2b}%
$(\langle z \rangle +V_d \, \tau)/{\langle z_0 \rangle}$ when
 $\sqrt{{\eta}/{g}}\leq\tau_p\leq\sqrt{{L}/{g}}$, ${V_d}/{u'}>1$
for
$Fr_d >1$ (case E\ref{sec:43} in table~\ref{tableIII}).
}
\end{figure}
Fig.~\ref{fig:grad2b} shows that when ${L}/{\tau_pg} < 1/\mathcal{N}$, the inertial
particles still fall down with the velocity $V_d$
(results are identical for cases D\ref{sec:43} and F\ref{sec:43} in table~\ref{tableIII}.)
as it was for the case when
$L/\tau_{p}g > 1/\mathcal{N}$.
\\[2ex]
The variance of the vertical inertial particle position $\langle (\zeta_3-\langle
\zeta_3 \rangle)^2 \rangle$ can be calculated directly and is found to oscillate around a
constant which scales as $(u'\tau_p)^2$ as can be seen from Fig.~\ref{fig:fig107}.
%
\begin{figure}[h]
\includegraphics[width=7cm]{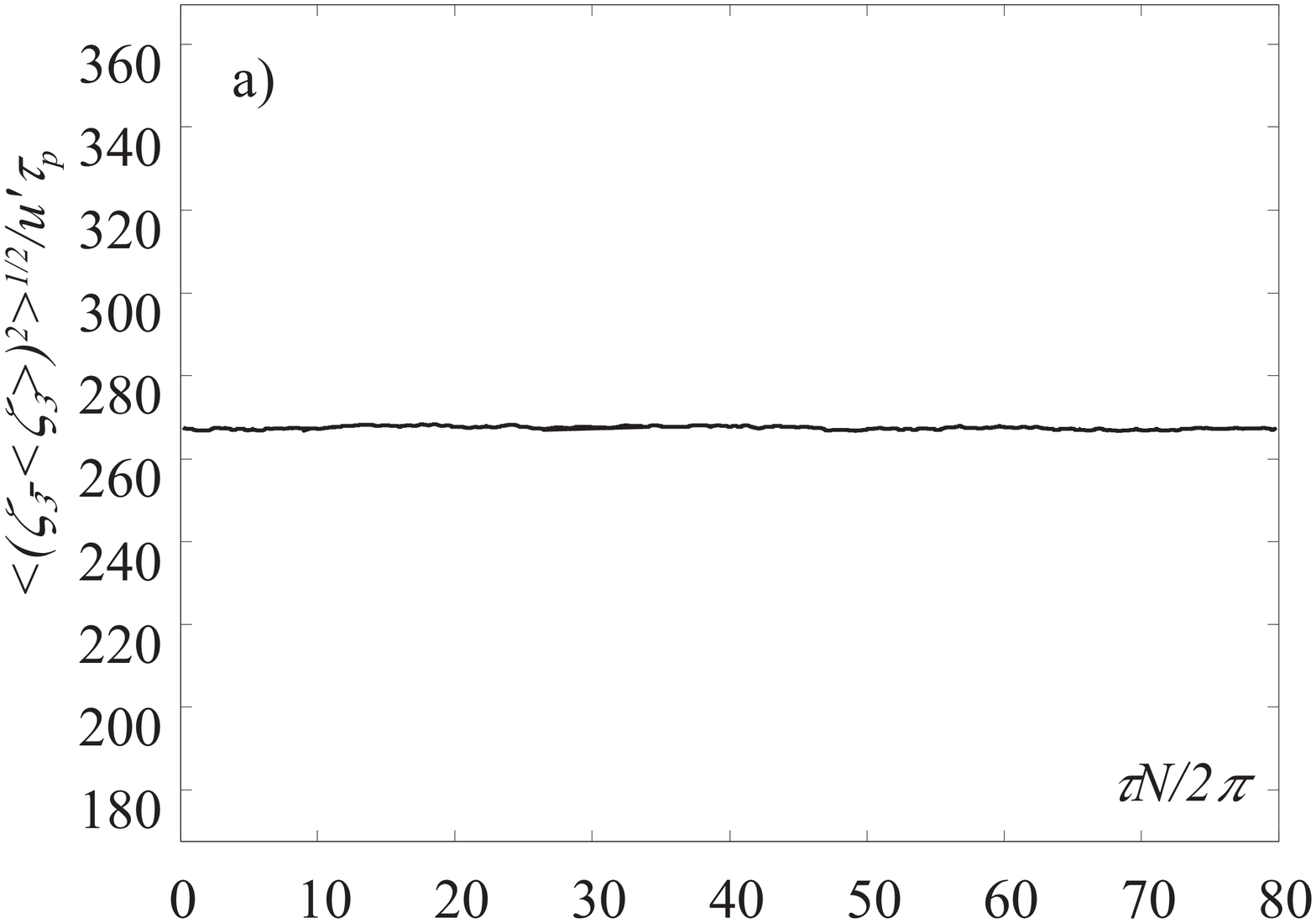}
\includegraphics[width=7cm]{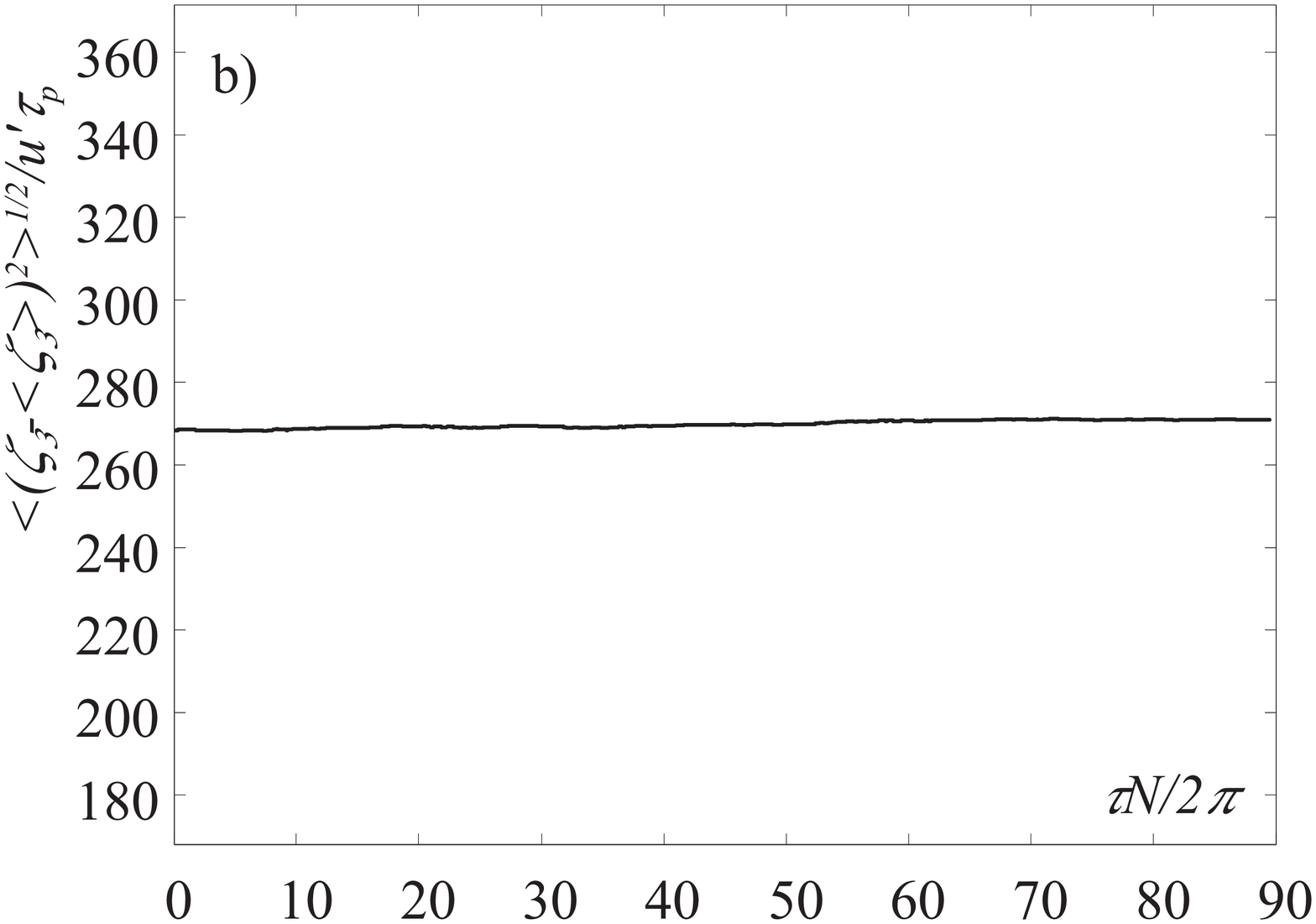}
\caption{%
\label{fig:fig107}%
$\sqrt{{<(\zeta_3-\langle \zeta_3 \rangle)^2>}}/{u'\tau_p}$ when
$\sqrt{{\eta}/{g}}<\tau_p<\sqrt{{L}/{g}}$ for
${L}/{\tau_pg}<{2\pi}/\mathcal{N}$ a) case E\ref{sec:43}, b) case H\ref{sec:43}.}
\end{figure}
\\[2ex]
The value of $\sqrt{{<(\zeta_3 -\langle
\zeta_3 \rangle)^2>}}/u'\tau_p$ remains the same when the buoyancy frequency
is increased, and is equal to 267:
\begin{equation}
\sqrt{<(\zeta_3 -\langle
\zeta_3 \rangle)^2>}
\simeq 267 u'\tau_p + \mbox{oscillations}
.
\label{zsqrfrouddl}
\end{equation}
Cases F\ref{sec:43}, G\ref{sec:43}, H\ref{sec:43} and I\ref{sec:43} not shown here yield identical results.
So by contrast to the previous case $Fr_d <1$ which was still obeying the fluid particle pattern (\ref{fp3stra}), when $Fr_d > 1$ the rms of the vertical position fluctuation
is still capped but it now obeys a different scaling which is independent of $\mathcal{N}$ and scales instead with $\tau_p$.
\\[2ex]
In Fig.~\ref{fig:cor1}, $R(\tau)$ is plotted as a function of $\tau N / 2 \pi$ for cases D\ref{sec:43}, E\ref{sec:43} and F\ref{sec:43} in Table~\ref{tableIII},
that is cases for which $Fr_d > 1$.
%
\begin{figure}[h]
\includegraphics[width=7cm]{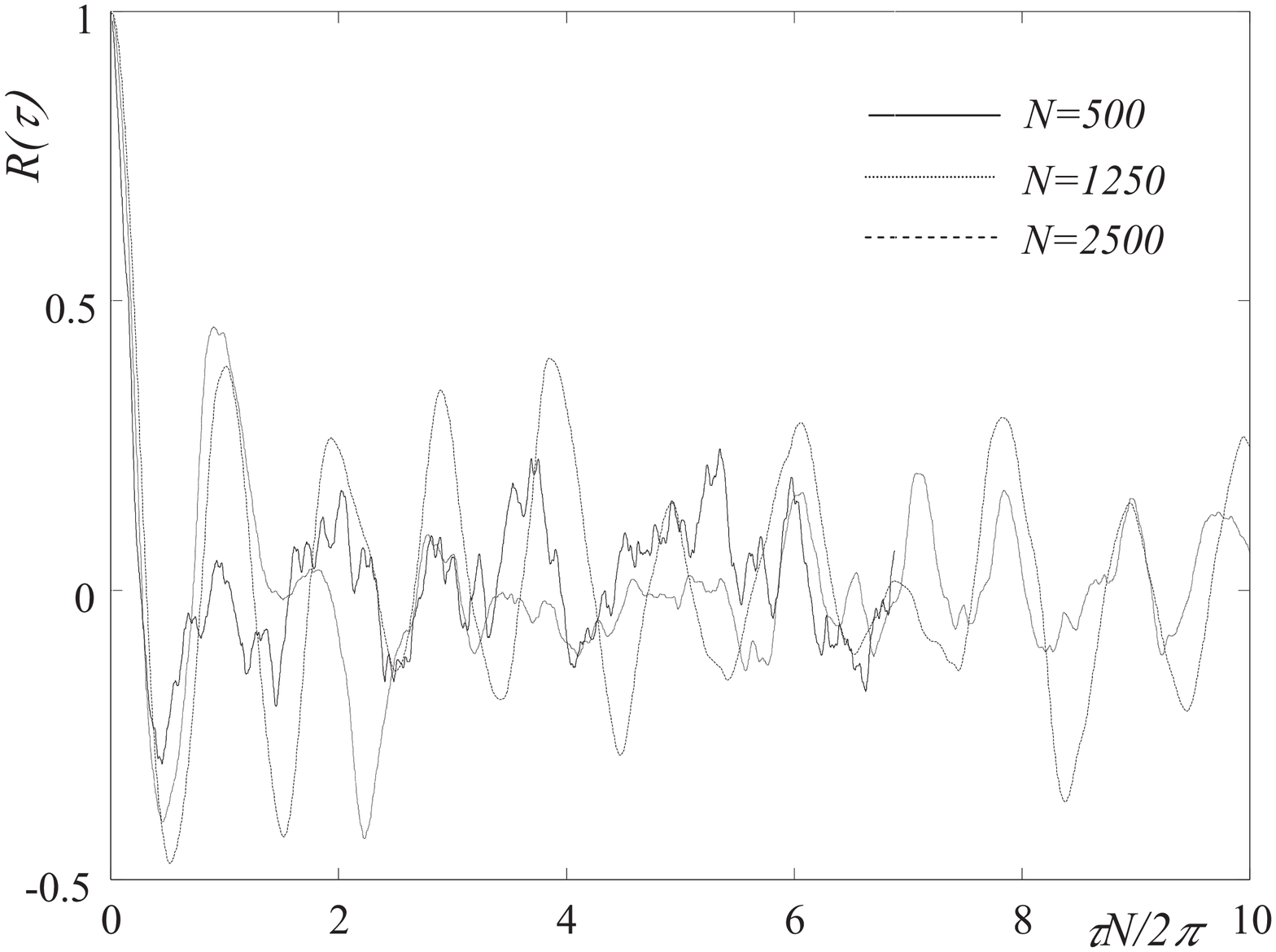}
\caption{%
\label{fig:cor1}
$R(\tau)$ as a function of $\tau N / 2 \pi$
when $Fr_d >1$. Cases D\ref{sec:43}, E\ref{sec:43} and F\ref{sec:43} in Table~\ref{tableIII}.}
\end{figure}
The autocorrelation is still oscillating around zero but the oscillations are more irregular and there is no clear $\mathcal{N}$ frequency as in Fig.~\ref{fig:cor1-1}. Case F\ref{sec:43}
for which $Fr_d=1.26$ is close to 1 is interesting, it shows a transitional behaviour with regular zero-crossings of frequency $\mathcal{N}$ but an irregular amplitude.
This is consistent with the previous findings (\ref{zsqrfrouddl}) that the dispersion scaling is independent of $\mathcal{N}$.
\\[2ex]
The particle's diffusivity can be obtained from
Taylor's relation (\ref{taylorlise}).
The normalised autocorrelation function $R(\tau)$ is integrated using
Simpson's 3/8 method, up to a time equal to many multiples of
$2\pi/\mathcal{N}$ and the integral is found to be much smaller than both
$\tau_p$ and $\sqrt{\eta/g}$ in all cases tried.
Table~\ref{tabletayl2} shows the values of the diffusivity obtained from Taylor's relation
for cases D\ref{sec:43}, E\ref{sec:43}, F\ref{sec:43} in table~\ref{tableIII}.
\begin{table}[h]
\caption{\label{tabletayl2}Integration of $\int R(\tau) d \tau$ for $Fr_d >1$. }
\begin{tabular}{rrr}
\hline\noalign{\smallskip}
case & $\mathcal{N}$ & $\int R(\tau) d \tau$
\\
\noalign{\smallskip}\hline\noalign{\smallskip}
D\ref{sec:43} & 500 & $4.61\times10^{-7}$
\\
E\ref{sec:43} & 1250 & $1.7\times 10^{-6}$
\\
F\ref{sec:43} & 2500 & $1.4\times 10^{-6}$
\\
\noalign{\smallskip}\hline
\end{tabular}
\end{table}
Though the oscillations shown in Fig.~\ref{fig:cor1} are irregular and completely different in nature when compared to the classical gravity-wave effect in Fig.~\ref{fig:cor1-1},
the integration is close to 0 for all the cases and we can conclude that there is no diffusivity and the vertical displacement variance is constant.
\\[2ex]
In conclusion, in this regime where $\sqrt{\eta/g} < \tau_p < \sqrt{L/g}$ the vertical diffusivity is zero and $\langle (\zeta_3-\langle\zeta_3\rangle)^2\rangle^{1/2}$ oscillates around a constant. This constant is proportional to $u'$ and to a time-scale which is different according to whether $Fr_d$ is smaller or larger than 1.
When $Fr_d<1$, $\langle (\zeta_3-\langle\zeta_3\rangle)^2\rangle^{1/2} \simeq 1.22 u'/\mathcal{N} + oscillations$. In this case, $1/\mathcal{N}<L/V_d$ and the autocorrelation function's oscillations are therefore dominated by buoyancy. Hence the time-scale controlling $\langle (\zeta_3-\langle\zeta_3\rangle)^2\rangle^{1/2}$ is $1/\mathcal{N}$.
However, when $Fr_d >1$, $\langle (\zeta_3-\langle\zeta_3\rangle)^2\rangle^{1/2} \simeq 267 u'\tau_p + oscillations$.
As described in \cite{Nicolleau-Vassilicos-2000} the plateau level is fixed by the end of the ballistic regime:
\[
\langle(x_3(t) - x_3(t_0))^2\rangle^{1/2} \sim {u'} \tau_b
\]
where $\tau_b$ is the duration of the ballistic regime that is $\simeq 1/\mathcal{N}$ in the case of a fluid particle.
Here $\tau_b=\tau_p$ as can be seen from Fig.~\ref{fig:cor1} were all the cases shown have the same $\tau_p$ but different $\mathcal{N}$; clearly the
end of the ballistic regime which corresponds to the beginning of the negative loops is independent of $\mathcal{N}$.

\section{Third regime:
$\sqrt{{L}/{g}}\leq\tau_p\leq\tau_{\eta}$}
\label{sec:sec2}

The last regime we consider in this section is $\sqrt{{L}/{g}}\leq\tau_p \leq \tau_{\eta}$.
This regime is also one where $W>1$. However, $Fr_d$ cannot be smaller than 1 in this regime because
$\sqrt{L/g}<\tau_p$. Hence, $Fr_d>1$.
Furthermore, according to condition~\ref{cond1}:
\begin{equation}
\sqrt{\frac{\eta}{g}}
<
\sqrt{\frac{L}{g}}
\leq
\tau_p
\leq
\tau_{\eta}.
\end{equation}
In terms of the large-scale-gravity-effect Stokes number,
\begin{equation}
St_{g L} >1
.
\end{equation}
In this third time-relaxation regime, we find that inertial particles fall down with
a velocity $V_d$, as we we now show.
%
%
%
\begin{table}[h]
\caption{
\label{tableII}%
Cases considered in section~\ref{sec:sec2}.}
\begin{tabular}{lrlrrr}
\hline\noalign{\smallskip}
case & $\tau_p$ & $V_d /u'$ & $ \mathcal{N} $ & $2 \pi / \mathcal{N}$ & $Fr_d$
\\
\noalign{\smallskip}\hline\noalign{\smallskip}
A\ref{sec:sec2} & $0.002$ & $ 2.0 \times 10^3$ & $ 500 $ & 0.0126 & 25.2
\\
B\ref{sec:sec2} & $0.002$ & $ 2.0 \times 10^3 $ & $ 1250 $ & 0.0050 & 10
\\
C\ref{sec:sec2} & $0.002$ & $ 2.0 \times 10^3 $ & $ 2500 $ & 0.0025 & 5
\\
D\ref{sec:sec2} & $0.035$ & $ 3.5 \times 10^4 $ & $ 500 $ & 0.0126 & 441
\\
E\ref{sec:sec2} & $0.035$ & $ 3.5 \times 10^4 $ & $ 1250 $ & 0.0050 & 175
\\
F\ref{sec:sec2} & $0.035$ & $ 3.5 \times 10^4 $ & $ 2500 $ & 0.0025 & 87.5
\\
\noalign{\smallskip}\hline
\end{tabular}
\end{table}

Computations have been made for all the cases in table~\ref{tableII}, but we present only one typical case in the figures as the other cases yield the same conclusion. Our results are not dependent on the particular values of $\tau_p$ and $\mathcal{N}$ within the constraint of this regime.

\subsection{Average vertical position}

\begin{figure}[h]
\includegraphics[height=4cm]{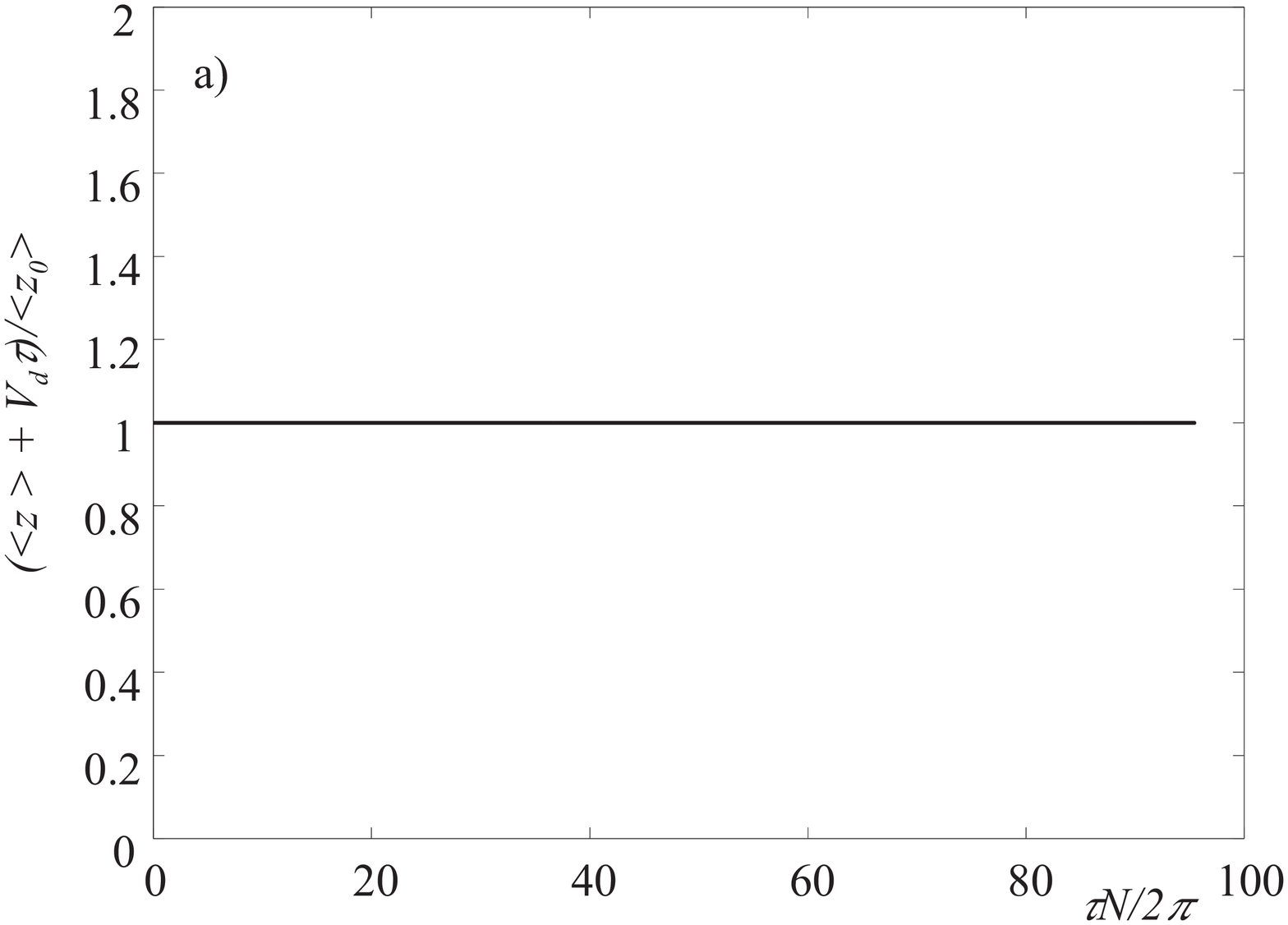}
\includegraphics[height=4cm]{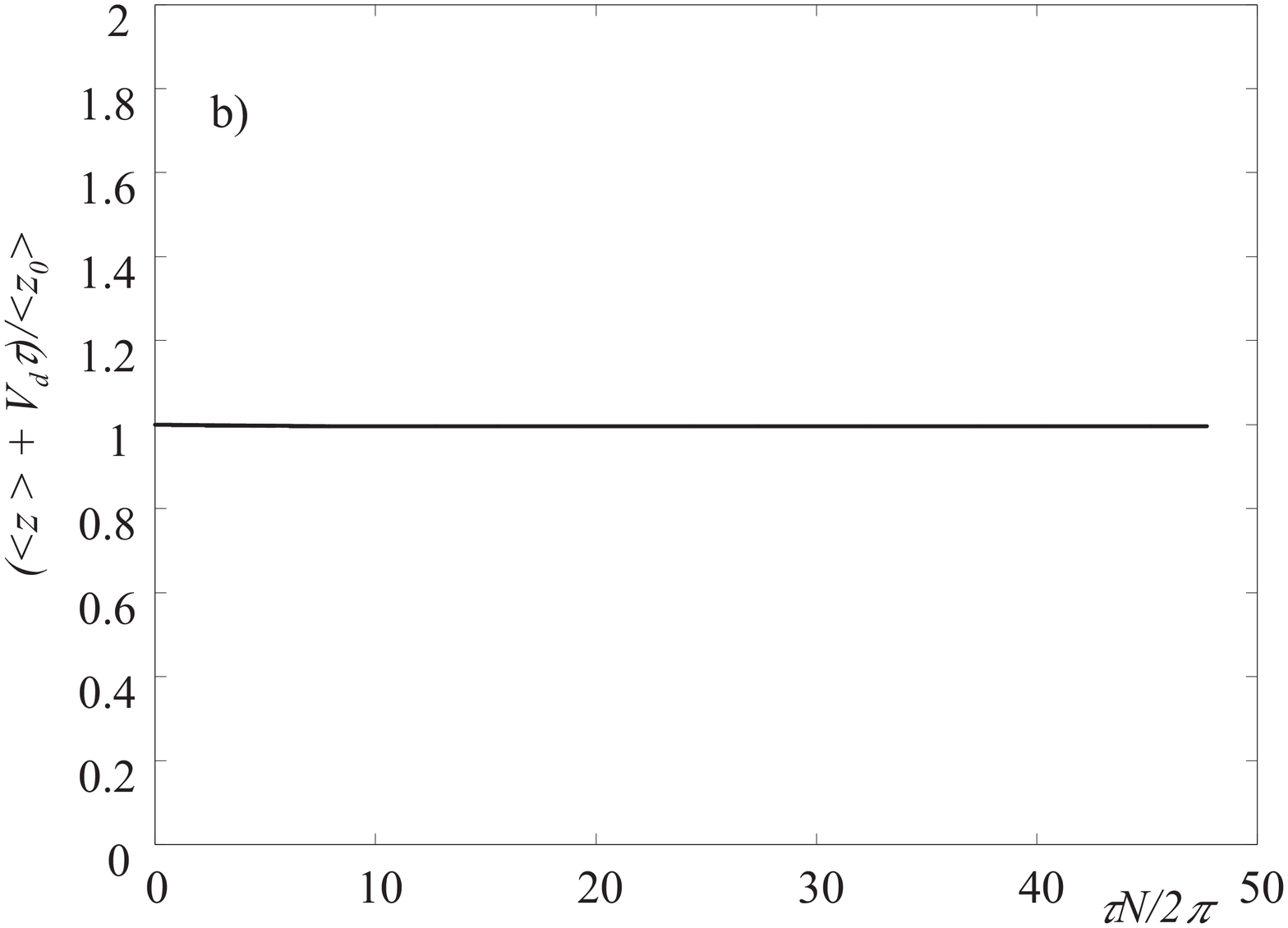}
\caption{%
\label{fig:grad3}%
$(\langle z \rangle + V_d \, \tau )/{\langle z_0 \rangle}$ as a function of $\tau \mathcal{N} / 2 \pi$ when
$W >1$ and
a) case A\ref{sec:sec2}
$\sqrt{{L}/{g}} < \tau_p < {2\pi}/\mathcal{N}$,
b) case D\ref{sec:sec2}
$\sqrt{{2\pi}/\mathcal{N}}\leq\tau_p\leq\tau_{\eta}$.}
\end{figure}

Fig.~\ref{fig:grad3}a shows
$(\langle z \rangle + V_d \, \tau)/{\langle z_0 \rangle}$ as a function of $\tau \mathcal{N} / 2 \pi$ when
$\sqrt{{L}/{g}} < \tau_p < {2\pi}/\mathcal{N}$ and
${V_d}/{u'}$ larger than 1.
Fig.~\ref{fig:grad3}b shows the same function as in Fig.~\ref{fig:grad3}a but this time for
$\sqrt{{2\pi}/\mathcal{N}}\leq \tau_p\leq\tau_{\eta}$.
In both cases $(\langle z \rangle + V_d \, \tau)/{\langle z_0 \rangle} \simeq 1$
and we can conclude that when $\sqrt{{L}/{g}}\leq\tau_p\leq\tau_{\eta}$
\[
\langle z \rangle = \langle z_0 \rangle - V_d \tau
.
\]
%

\subsection{Variance of the vertical displacement}

\begin{figure}[h]
\includegraphics[width=15cm]{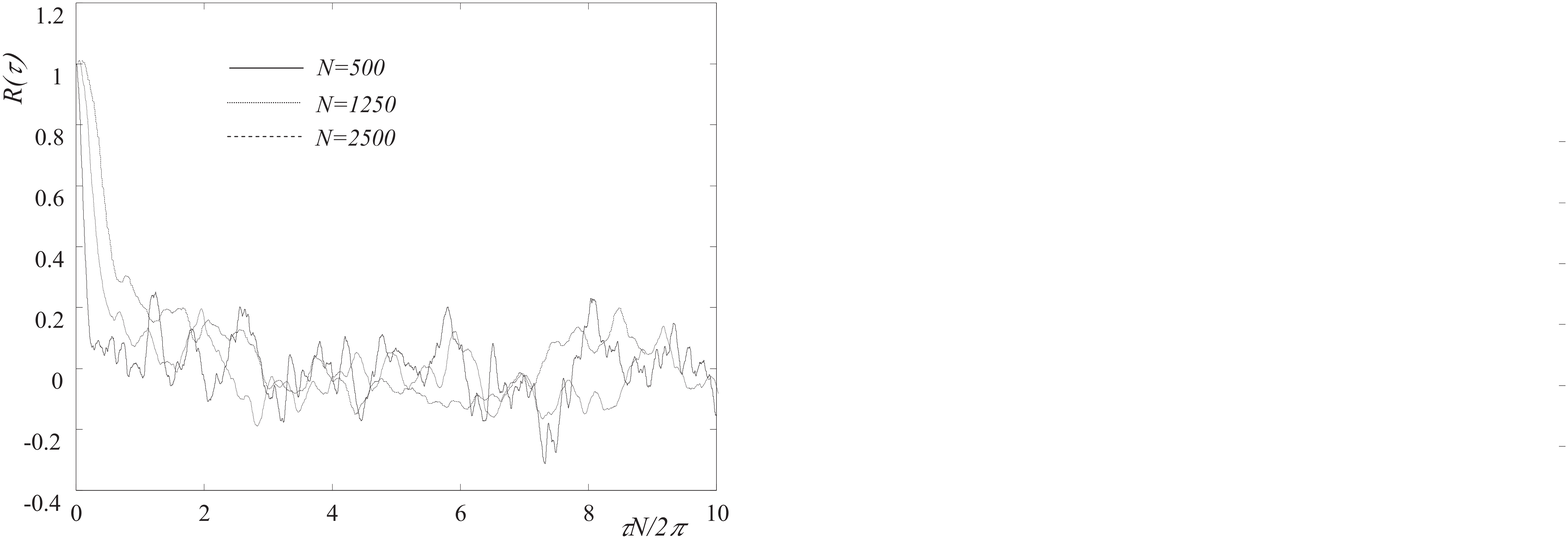}
\caption{%
\label{fig:cor2}
$R(\tau)$ as a function of $\tau N / 2 \pi$ for $\sqrt{{L}/{g}}<\tau_p<{2\pi}/\mathcal{N}$.}
\end{figure}

\begin{figure}[h]
\includegraphics[width=7cm]{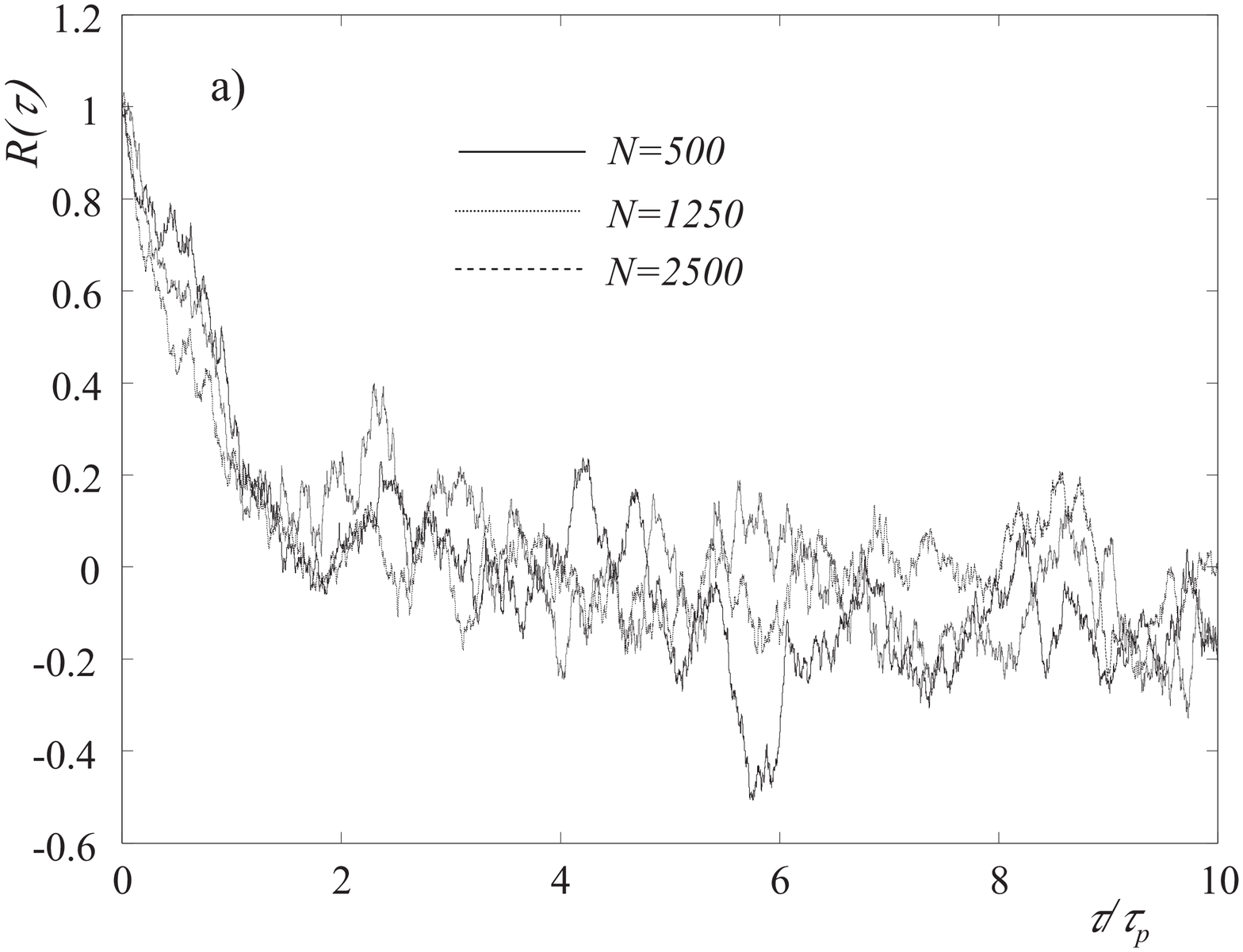}
\includegraphics[width=7cm]{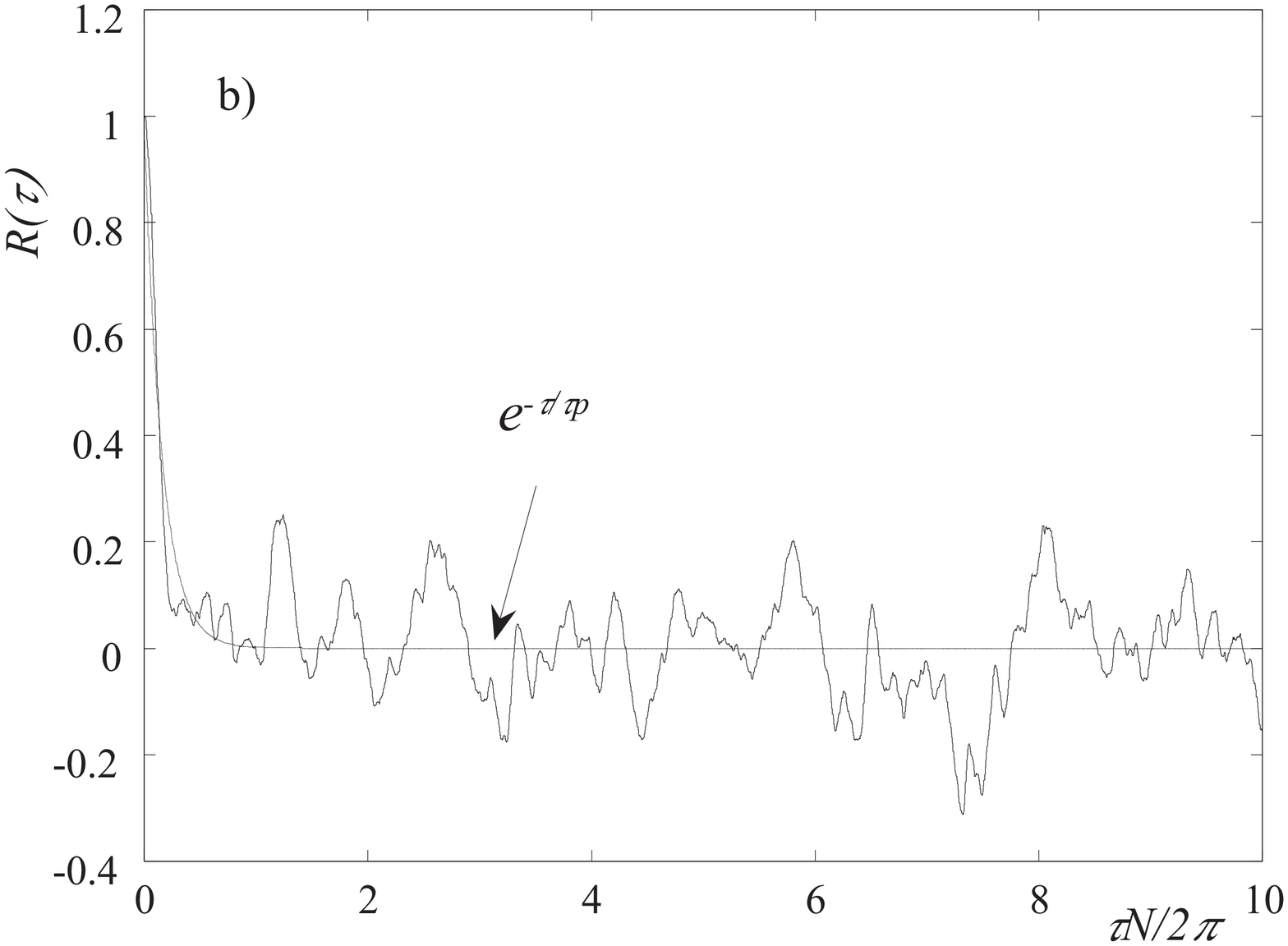}
\caption{%
\label{fig:cor2b}
a) $R(\tau)$ as a function of $\tau / \tau_p$ for ${2\pi}/\mathcal{N}<\tau_p<\tau_{\eta}$.
b) $R(\tau)$ as a function of $\tau N / 2 \pi$ for case A\ref{sec:sec2} solid line and $e^{-\tau/\tau_p}$ for that case dash line.}
\end{figure}

\begin{figure}[h]
\includegraphics[width=7cm]{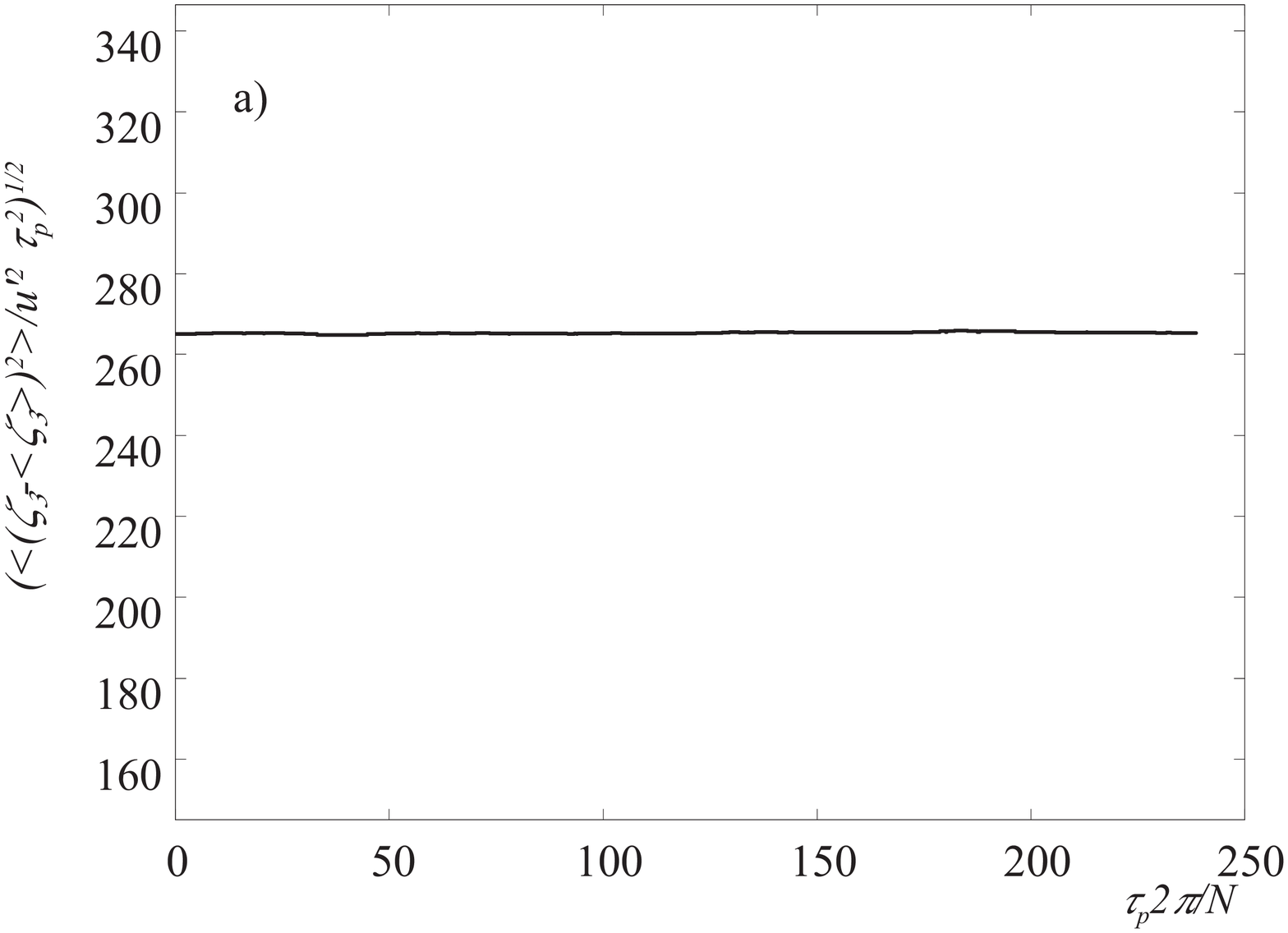}
\includegraphics[width=7cm]{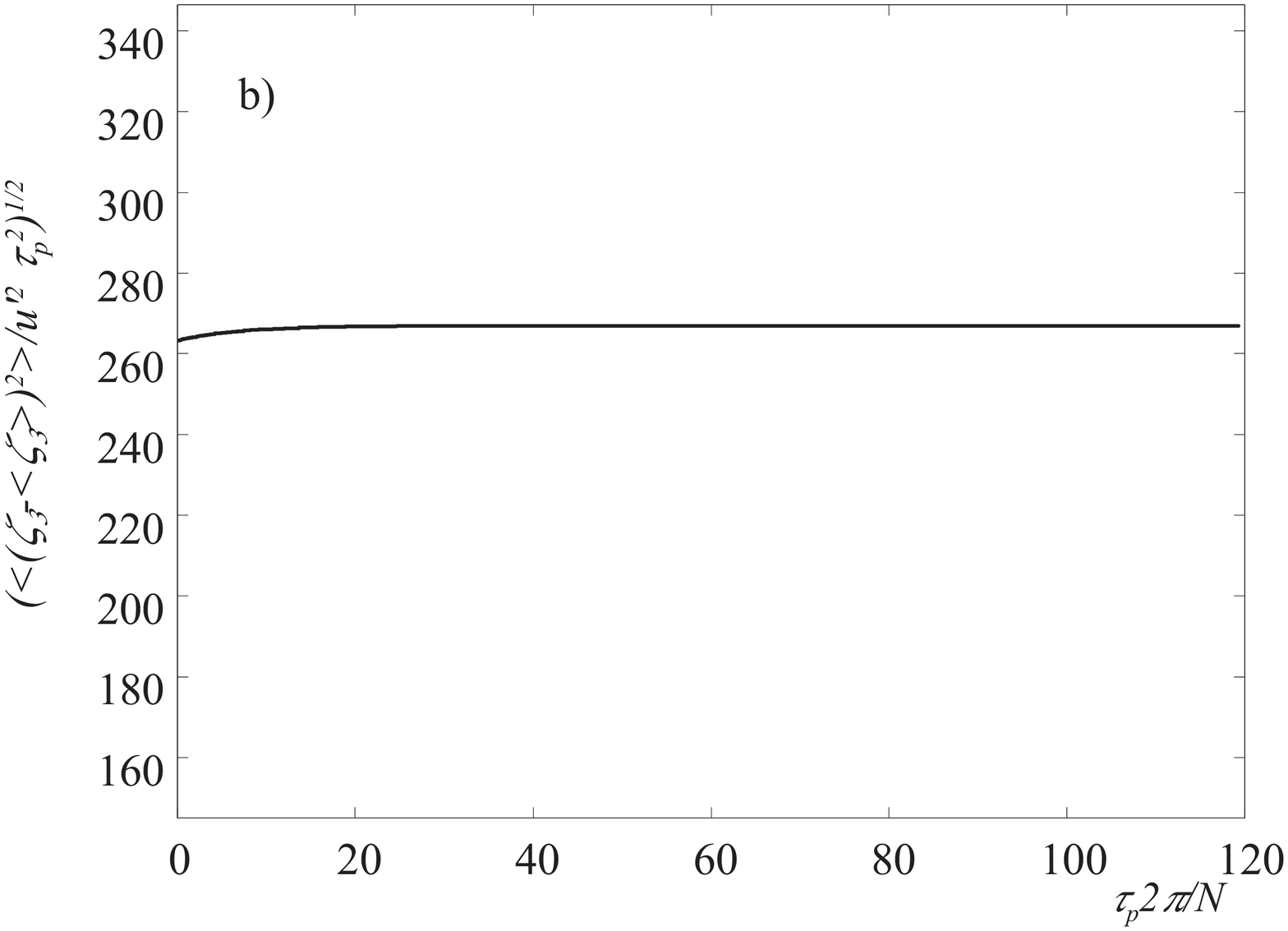}
\caption{
\label{fig:varz2}%
$\sqrt{{\langle (\zeta_3-\langle \zeta_3 \rangle)^2 \rangle}/{(u'\tau_p)^2}}$:
a)
$\sqrt{{L}/{g}}<\tau_p<{2\pi}/\mathcal{N}$, case B\ref{sec:sec2};
b)
${2\pi}/\mathcal{N}<\tau_p<\tau_{\eta}$, case E\ref{sec:sec2}.}
\end{figure}

We can first get an idea of the displacement variance by looking at the velocity auto\-correlation.
In Fig.~\ref{fig:cor2} the autocorrelation
function is plotted for $\tau_p=0.002$ that is $\sqrt{{L}/{g}}<\tau_p<{2\pi}/\mathcal{N}$ which corresponds to case A\ref{sec:sec2}, B\ref{sec:sec2} and C\ref{sec:sec2}
in Table~\ref{tableII}.
Clearly the relevant time scale is not $\tau N / 2\pi$. In Fig.~\ref{fig:cor2b}a we use a different normalisation in time namely $\tau / \tau_p$
for the cases $\tau_p=0.035$ that is ${2\pi}/\mathcal{N}<\tau_p<\tau_{\eta}$ which corresponds to case D\ref{sec:sec2}, E\ref{sec:sec2} and F\ref{sec:sec2} in Table~\ref{tableII}.
All the curves collapse (the same collapse would be observed for the cases $\sqrt{{L}/{g}}<\tau_p<{2\pi}/\mathcal{N}$). So clearly the relevant time scale is now $\tau_p$, and this must be because $Fr_d >1$.
This is confirmed in Fig.~\ref{fig:cor2b}b where we superimpose the curve $e^{-\tau/\tau_p}$ onto case A\ref{sec:sec2} from Fig.~\ref{fig:cor2}. It shows clearly that for $\sqrt{{L}/{g}}\leq\tau_p\leq\tau_{\eta}$ the autocorrelation is oscillating around the vanishing exponential $e^{-\tau/\tau_p}$.
\\[2ex]
The integration of the autocorrelation
function is carried out with Simpson's 3/8 rule, the
results are shown in table~\ref{tabtaylor3}. This integration is found to be
very close to 0 at all time scales.
\begin{table}[h]
\caption{%
\label{tabtaylor3}
Note that $\sqrt{\eta/g}=0.0001$.}
\begin{tabular}{lrr}
\hline\noalign{\smallskip}
Case & $\mathcal{N}$ & Taylor's relation
\\
\noalign{\smallskip}\hline\noalign{\smallskip}
$\tau_p < {2\pi /\mathcal{N}}$ & 500 & $7.60\times 10^{-8}$
\\
$\tau_p < {2\pi /\mathcal{N}}$ & 1250 & $4.17\times 10^{-7}$
\\
$\tau_p < {2\pi /\mathcal{N}}$ & 2500 & $2.38\times 10^{-7}$
\\
$\tau_p > {2\pi /\mathcal{N}}$ & 500 & $3.20\times 10^{-7}$
\\
$\tau_p > {2\pi /\mathcal{N}}$ & 1250 & $7.60\times 10^{-9}$
\\
$\tau_p > {2\pi /\mathcal{N}}$ & 2500 & $2.70\times 10^{-7}$
\\
\noalign{\smallskip}\hline
\end{tabular}
\end{table}
This suggests that there is no vertical diffusivity
in this relaxation time regime. But as the autocorrelation's decorrelation is controlled by $e^{-\tau/\tau_p}$
provided there are significant oscillations which causes the integral of $R(\tau)$ to vanish
 (as shown in Fig.~\ref{fig:cor2b}), we can expect the vertical variance to scale with $\tau_p$ rather than $\mathcal{N}$.
\\[2ex]
The variance of the vertical inertial particle position
$\langle (z-\langle z\rangle)^2 \rangle$ is studied in
Fig.~\ref{fig:varz2} as a function of time for the two cases
$\sqrt{{L}/{g}}<\tau_p<{2\pi}/\mathcal{N}$ and $\sqrt{{2\pi}/\mathcal{N}}<\tau_p<\tau_{\eta}$
corresponding respectively to B\ref{sec:sec2} and E\ref{sec:sec2} in table~\ref{tableII}.
This ratio is clearly a constant, more precisely we retrieve relation~\ref{zsqrfrouddl}:
\[
\sqrt{\langle (\zeta_3 -\langle \zeta_3 \rangle)^2 \rangle} \simeq 267 u' \tau_p
\]
that was observed in section~\ref{weak4}.
This value 267 remains the
same irrespective of the value of the buoyancy frequency.

If we include the results from section~\ref{weak4}, we can conclude as at the end of section \ref{sec:43} that
the critical condition for whether the vertical diffusion of inertial particles
is dominated by the falling effect of gravity or the oscillatory
effect of gravity waves is whether
${L}/{\tau_pg} < {2\pi}/\mathcal{N}$ or $ > {2\pi}/\mathcal{N}$. So the key parameter is $Fr_d$.
When $Fr_d >1$ relation~\ref{zsqrfrouddl} holds.
\begin{figure}[h]
\includegraphics[width=7cm]{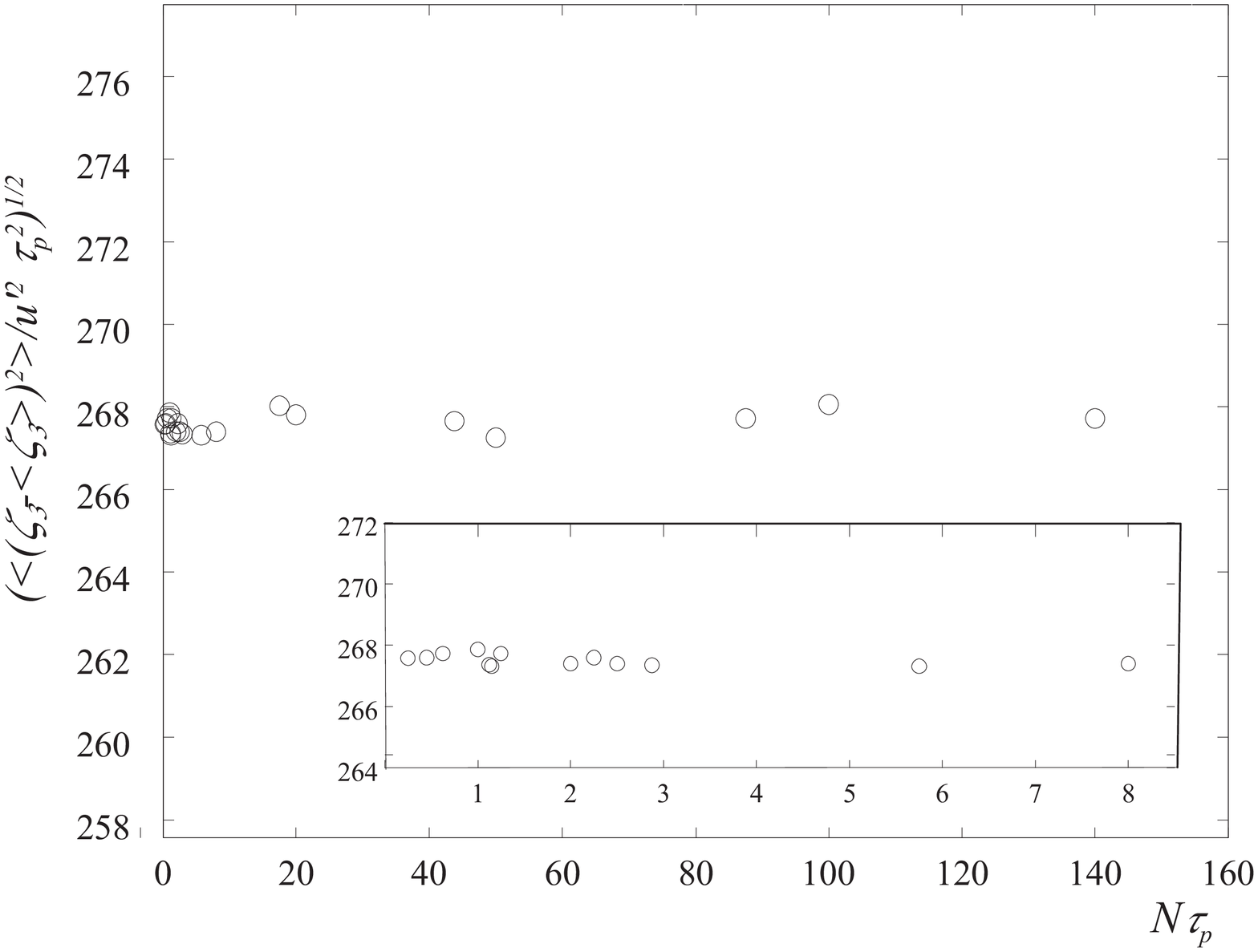}
\caption{%
\label{fig:varcf3}%
Coefficient $\sqrt{\langle(\zeta_3-\langle \zeta_3 \rangle)^2\rangle}/u'\tau_p$ as a function of
$\mathcal{N}\tau_p$, for
$Fr_d > 1$
The insert graph is a magnification for the range
$0 \le \mathcal{N}\tau_p \le 9$.}
\end{figure}

In Fig.~\ref{fig:varcf3} we plot the asymptotic value of $\sqrt{\langle (\zeta_3-\langle \zeta_3 \rangle)^2 \rangle}/u'\tau_p$ as a function of $\mathcal{N}\tau_p$ when $Fr_d >1$
for different values of the buoyancy frequency and of $\tau_p$.
Namely, $\mathcal{N}=500$, 1250 and 2500 and $\tau_p=$0.0005,
0.00009, 0.002, 0.0023, 0.035 and 0.04. In this figure, the largest drift characteristic time
${L}/V_d$ is 0.002 for $\tau_p=$0.0005 and the smallest buoyancy characteristic time
${2\pi}/\mathcal{N}$ is 0.0025 for $\mathcal{N}=2500$.
In all cases we observe the scaling
\[
\langle (\zeta_3-\langle \zeta_3\rangle)^2 \rangle^{1/2} \simeq 267u'\tau_p
.
\]
When
$\sqrt{{L}/{g}}<\tau_p$, then ${L}/V_d$ is always
smaller than both $\tau_p$ and the stratification time scale ${2\pi}/\mathcal{N}$.
It may be surprising that the conclusion at end of \ref{sec:43} still holds because $L/V_d$ is now smaller than $\tau_p$.
However, $L/V_d$ is the time scale which controls the average fall and $\tau_p$ is the time scale which controls the decay of $R(\tau)$ when $Fr_d > 1$ as is the case here.
Hence, we expect
\[
<(\zeta_3-\langle \zeta_3 \rangle)^2>^{1/2}\sim u'\times \mbox{(ballistic time of inertial particles)} \sim u'\tau_p
.
\]
This
is indeed what is observed in Fig.~\ref{fig:varz2}. This explains only the scaling
which is characteristics of a decorrelating time $\tau_p$ as can be seen in Fig.~\ref{fig:cor2b}b. However, though this displacement is decorrelating apparently without
an identified regular frequency the diffusivity is 0 and the dispersion is capped (Fig.~\ref{fig:varz2}). From Fig.~\ref{fig:cor2b}b we can see that the correlation is oscillating with loops
above and below the mean trend $e^{-\tau/\tau_p}$ ensuring the capping of the vertical dispersion. So that we can conclude that, in our KS field,
the displacement retains the memory of being in a stratified flow even for $Fr_d>1$.

\section{Conclusion}

We use a synthetic model of turbulence (KS)
to study the vertical dispersion of heavy particles in stratified flows. The model we use limits the range of stratifications we can study:
the underlying Boussinesq approximation and RDT validity impose:
\begin{equation}
\sqrt{\frac{\eta}{g}}
<
\sqrt{\frac{L}{g}}
<
\frac{2\pi}{\mathcal{N}}
<
\tau_{\eta}
<
\frac{L}{u'}
\label{timescalebr}
\end{equation}

\begin{figure}[h]
\epsfig{file=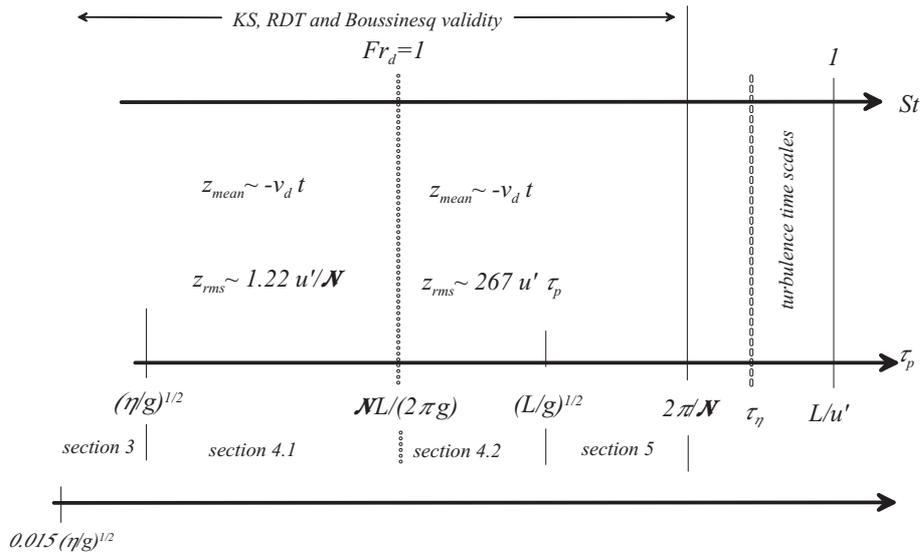, width=12cm}
\caption{\label{fig:schematic}%
Schematic of the different
relaxation time regimes.}
\end{figure}

The first conclusion valid for all the regimes we have studied is that
though the particle with inertia falls with a terminal velocity $V_d$ as it would in a turbulence without stratification, the variance of its fluctuation $\zeta$
is capped in the vertical direction as it would be for a fluid particle in a stratified turbulence
and there is no vertical diffusivity.
However, the value of the plateau is not always that of
the fluid particle depending on the value of the Froude number $Fr_d$.
\\[2ex]
The five different time scales identified in (\ref{timescalebr}), are shown in Fig.~\ref{fig:schematic}.
%
\\[2ex]
The critical
condition for the vertical diffusion of inertial particles
to be dominated by the falling effect of gravity or the oscillatory
effect of gravity waves is whether
${L}/{\tau_pg} < {2\pi}/\mathcal{N}$ or $ > {2\pi}/\mathcal{N}$. So the relevant non dimensional number is
\[
Fr_d=\frac{2 \pi / \mathcal{N}}{L/V_d} = W \, Fr
\]
where we have introduced the (usual large scale) Froude number
$Fr= 2 \pi {u'}/{L\mathcal{N}}$ and the drift parameter $W = V_d / u'$.
%
%
Results can be summarized as follows:
\begin{itemize}
\item
When $Fr_d < 1$, whether $St_{g \eta} = \tau_p \sqrt{(g/\eta)}$ is larger or smaller than 1,
the vertical position of inertial particles is such that
\begin{equation}
\langle z \rangle \simeq \langle z_0 \rangle - V_d \, \tau
\label{eqn:summ1}
\end{equation}
\begin{equation}
\langle (z-z_0 - \langle z-z_0\rangle)^2 \rangle^{1/2}\simeq 1.22 \frac{u'}{\mathcal{N}} + \mbox{ oscillations}
\label{eqn:summ3}
\end{equation}
In this relaxation
time regime, the plateau for the rms of the displacement fluctuation is
that found for fluid particles in stably stratified turbulence.
\\[2ex]
This is valid for drift parameters
$W = \tau_pg/u'$ smaller or larger than 1 and any value of $St_{g \eta}$.
\item
When $Fr_d > 1$ (which implies $\sqrt{{\eta}/{g}} < \tau_p$) and whether $\tau_p < \sqrt{{L}/{g}}$ or $\sqrt{{L}/{g}} < \tau_p < \tau_{\eta} $
the
inertial particles still fall down with a gradient $V_d$ and
\begin{equation}
\langle z \rangle \simeq \langle z_0 \rangle - V_d \, \tau,
\end{equation}
KS results show that the variance of the fluctuation of the vertical displacement is constant in time demonstrating that there is no vertical
diffusivity. However, in this regime the value for the plateau is not that found for fluid particles. In particular, it is independent of
$\mathcal{N}$ but a only a function of turbulence intensity ($u'$) and $\tau_p$. The exact relation we found with KS is
\begin{equation}
\langle (z-z_0-\langle z-z_0 \rangle)^2 \rangle^{{1}/{2}}\simeq 267 \, u'\tau_p
.
\end{equation}
\end{itemize}
%
This is all the more surprising since as discussed in \cite{Cambon-al-2004,Nicolleau-Yu-2007} 
KS do not possess the Eulerian spatial `structuration' found in DNS \cite{Liechtenstein-al-2005} and therefore
the vertical dispersion capping in KS is only controlled by the Lagrangian correlations. These latter are of course controlled in turn by the two-time Eulerian correlations as some basic analytical integrations would show \cite{VanHarrenPhD}.
Keeping a plateau while having a decorrelating time-scale based on the turbulence
time-scale could be easier to understand if there was a two-time and a two-point structuration
imposing different physical times. Though from previous studies on fluid particle there is no
indication that the Eulerian space correlation plays a role in the Lagrangian plateau and scaling, it would be interesting to generalise this result to particles with inertia.
Whether the Eulerian space structurations found in DNS enhance or annihilate the new regime predicted by KS for $Fr_d >1$ remains an open question.
\\[2ex]
Furthermore, KS does not account for the sweeping effect which
states that energy containing turbulent eddies advect small scale
dissipative turbulent eddies \cite{Tennekes-1975,Nicolleau-Nowakowski-2011}. It has been found
in isotropic homogeneous turbulence that heavy particles stick and
move with regions where fluid acceleration is zero, or very small
\cite{Chen-al-2006}. This sweeping effect may alter our
proposed Stokes numbers but gravity and stratification may also
alter the effect found by \cite{Chen-al-2006}.

\begin{acknowledgements}
This work was supported by the Engineering and Physical
Sciences Research Council through the UK Turbulence
Consortium (Grant No. EP/G069581/1). F. Nicolleau also gratefully acknowledges support from the Leverhulme Trust (Grant No F/00 118/AZ).
\end{acknowledgements}


\bibliographystyle{spmpsci}


\end{document}